\documentclass{aa}  
\usepackage{array}
\usepackage{graphicx}
\usepackage{amsmath}
\usepackage{natbib}
\usepackage[version=3]{mhchem}
\usepackage{txfonts}
\titlerunning{{\hhhp} in the CMZ}

\def\hh{H$_2$}
\def\hhp{H$_2^+$}
\def\hhhp{H$_3^+$}

\def\nh{$n_{\textrm{H}}$}

\def\cmd{cm$^{-2}$}
\def\cmt{cm$^{-3}$}
\def\av{$A_\textrm{V}$}
\def\kms{km s$^{-1}$}

\def\hm{\textrm{H}}
\def\hhm{\textrm{H}_2}

\def\hhhpm{\textrm{H}_3^+}

\def\vlsr{v$_{LSR}$}

\begin{document}

\title{Physical conditions in the central molecular zone inferred by H$_3^+$}

\author{
Franck Le Petit\inst{1}
\and
Maxime Ruaud\inst{2,3}
\and
Emeric Bron\inst{1}
\and
Benjamin Godard\inst{1}
\and
Evelyne Roueff\inst{1}
\and
David Languignon\inst{1}
\and
Jacques Le Bourlot\inst{1,4}
}

\institute{ LERMA, Observatoire de Paris, PSL Research University, CNRS, UMR8112, F-92190 Meudon, France\\
Place J. Janssen, 92190 Meudon, France\\
\email{Franck.LePetit@obspm.fr}
\and
Univ. Bordeaux, LAB, UMR 5804, 33270 Floirac, France
\and
CNRS, LAB, UMR 5804, F-33270 Floirac, France
\and
Univ. Paris-Diderot, Paris 7\\
}

\date{Working draft}

\abstract
{The H$_3^+$ molecule has been detected in many lines of sight within the central molecular zone (CMZ) with exceptionally large column densities and unusual excitation properties compared to diffuse local clouds. The 
 detection of the (3,3) metastable level has been suggested to be the signature of warm and diffuse gas in the CMZ.
}
{We aim to determine the physical conditions and processes in the CMZ that explain the ubiquitous properties of H$_3^+$ in this medium and to constrain the value of the cosmic-ray  ionization rate.}
{We use the Meudon PDR code in which H$_3^+$ excitation has been implemented. We re-examine the relationship between the column density of H$_3^+$ and the cosmic-ray ionization rate, $\zeta$, up to large values of $\zeta$ in the frame of this full chemical model. We study the impact of the various mechanisms that can excite \hhhp~in its metastable state. We produce  grids of PDR models exploring different parameters ($\zeta$, size of clouds, metallicity) and infer the physical conditions that best match the observations toward ten lines of sight in the CMZ. For one of them, Herschel observations of HF, OH$^+$, H$_2$O$^+,$ and H$_3$O$^+$ can be used as additional constraints. We check that the results found for H$_3^+$ also account for the observations of these molecules.}
{We find that the linear relationship between $N(\textrm{H}_3^+)$ and $\zeta$ only holds up to a certain value of the cosmic-ray ionization rate, which  depends on the proton density. A value $\zeta \sim 1 - 11 \times 10^{-14}$ s$^{-1}$ explains both the large observed \hhhp~column density and its excitation in the metastable level (3,3). This $\zeta$ value agrees with that derived from synchrotron emission and Fe K$\alpha$ line. It also reproduces $N$(OH$^+$), $N$(H$_2$O$^+$) and $N$(H$_3$O$^+$) detected toward Sgr B2(N). We confirm that the CMZ probed by H$_3^+$ is diffuse, \nh~$\lesssim$ 100 \cmt~and warm, T $\sim$ 212-505 K. This warm medium is due to cosmic-ray heating. We also find that the diffuse component probed by \hhhp~must fill a large fraction of the CMZ. Finally, we suggest the warm gas in the CMZ enables efficient H$_2$ formation via chemisorption sites as in PDRs. This contributes to enhance the abundance of \hhhp~in this high cosmic-ray flux environment.}
{}

\keywords{astrochemistry - molecular processes - ISM:molecules}

\maketitle

\section{Introduction}
\label{Sec:intro}

\hhhp~has been observed in a variety of environments: dense clouds \citep{Geballe96, McCall99, Brittain04, Gibb10}, diffuse clouds \citep{McCall98, Geballe99, McCall02, McCall03, Indriolo07, Indriolo12} and even the nucleus of an extragalactic source \citep{Geballe06}. A comprehensive review has been recently published \citep{Oka13}. The cosmic-ray ionization rate can be inferred using  \hhhp~\citep{McCall02, McCall03, Pap2004}. \cite{Indriolo07} and \cite{Indriolo12} present an exhaustive study of \hhhp~observations and conclude that, in local diffuse clouds, the cosmic-ray ionization rate\footnote{We note $\zeta$ the cosmic-ray ionization rate of H$_2$, expressed in s$^{-1}$, and $\zeta_1$ the one of H with $\zeta_1$  $\sim$ 0.5 $\times$ $\zeta$.} $\zeta$ presents variations with a mean value equals to $3.5\times10^{-16}$ s$^{-1}$. The first detection of \hhhp~in the central molecular zone (CMZ) \citep{Geballe99} showed a surprisingly large column density. Subsequent detections in the CMZ \citep{Goto02, Oka05, Goto08, Geballe10, Goto11, Goto13, Goto14} revealed that, in a radius of at least 100 pc around Sgr A*, \hhhp~presents not only large column densities but also a peculiar excitation since it is observable in its (3,3) metastable level lying at 361 K above the ground rotational state (see Fig. \ref{Fig:LevelsH3p}). \cite{Oka04} established that $N$(3,3)/$N$(1,1) and $N$(3,3)/$N$(2,2) ratios  can be used to infer gas density and temperature. They demonstrated that a large fraction of the CMZ must be neutral diffuse warm gas with \nh~ < 100 cm$^{-3}$ and $T \sim 200-300$ K \citep{Oka04, Oka05, Goto08}. Two important questions arise from these unusual observations: 1) what is the cosmic-ray ionization rate required to explain these large amounts of \hhhp; and 2) what is the heating source responsible for the warm medium. To answer the first question, \cite{Oka05} and \cite{Goto08} used a simple analytic relation between $N(\textrm{H}_3^+)$ and $\zeta$. \cite{Oka05} concluded that $\zeta \sim (2-7) \times 10^{-15}$ s$^{-1}$, i.e., a cosmic-ray ionization rate typically ten times higher than in local diffuse clouds. More recently, \cite{Yusef13} used synchrotron emission and Fe I K$\alpha$ line to constrain $\zeta$. They found $\zeta_1 \sim 5\times 10^{-15}$ s$^{-1}$ with synchrotron emission and $10^{-14}$ s$^{-1}$ with the Fe I K$\alpha$ line. Finally, Herschel observations of OH$^+$, H$_2$O$^+$, and H$_3$O$^+$ in the CMZ suggest $\zeta_1 \sim 10^{-14}$ s$^{-1}$ toward Sgr B2(M) and (N) \citep{Indriolo15}.\\

In this paper, we use the Meudon PDR code\footnote{The Meudon PDR code is available at http://ism.obspm.fr} \citep{Pap2006} to study the properties of \hhhp~in a medium submitted to a high cosmic-ray ionizing flux and we propose a scenario that explains both the large \hhhp~column densities observed in the CMZ and its excitation in the (3,3) metastable  level. In Sect. \ref{Sec:ChemH3p}, we examine the relationship between $N(\textrm{H}_3^+)$ and the cosmic-ray ionization rate for a large range of $\zeta$ (from $10^{-17}$ to $10^{-12}$ s$^{-1}$) and various densities. We also study the impact of several processes on \hhhp~abundance, such as the recombination  of electrons on grains and H$_2$ formation mechanisms. In Sect. \ref{Sec:Excit}, we present an analysis of \hhhp~excitation based on new \hhhp~excitation collisional rates with \hh~computed by \cite{Roncero12}. We compare these new results to the pioneering study of \cite{Oka04}. We also study the possibility of \hhhp~excitation by nonthermal processes, such as IR pumping and excitation at formation. In Sect. \ref{Sec:CMZ} we present a scenario accounting for both $N(\textrm{H}_3^+)$ and \hhhp~excitation, as observed in the CMZ. We check that our results are consistent with other observational tracers of cosmic rays: OH$^+$, H$_2$O$^+$ and H$_3$O$^+$  absorptions in the CMZ, synchrotron, and Fe K$\alpha$ line emission. Finally, we compare the various heating and cooling mechanisms in the CMZ according to our model. Our conclusions are summarized in Sect. \ref{Sec:conclusion}.

\section{Abundance of H$_3^+$ in diffuse gas}
\label{Sec:ChemH3p}

Cosmic-ray ionization is the starting point of chemistries involving H, D, O, and N in diffuse and translucent clouds since these atoms cannot be photoionized by the interstellar radiation field (ISRF), see the review by \cite{Grenier15}. OH and HD have been recognized early \citep{Hartquist78,Federman96} as potential indicators of the cosmic-ray ionization rate $\zeta_1$ and a value of $\sim10^{-17}$ s$^{-1}$ was derived from the analysis of the column densities of OH and HD found in local diffuse clouds. Recently, \cite{Bialy15} studied the chemistry of OH for various $\zeta$/n$_\textrm{H}$ and low metallicities. Detections of \hhhp~toward various diffuse lines of sight \citep{McCall02, McCall03, Indriolo07, Indriolo12} lead to a significantly larger cosmic-ray ionization rate. This conclusion relies on a simple analytic relation between $N(\textrm{H}_3^+)$ and $\zeta$ derived from the chemical balance between formation and destruction of \hhhp~(see below). More recently, \cite{Indriolo15} used Herschel observations of OH$^+$, H$_2$O$^+$, and H$_3$O$^+$ in a sample of Galactic diffuse clouds and inferred a mean value of the cosmic-ray ionization rate in the local medium of $\zeta_1 \sim 1.78\times10^{-16}$ s$^{-1}$, using the same simple analytic expression based on chemical networks.\\

In this section, we revisit the relationship between $N(\textrm{H}_3^+)$ and $\zeta$ and extend it to large values of the cosmic-ray ionization rate, as those may exist in the Galactic center. We also highlight the indirect effect of two processes on $N(\textrm{H}_3^+)$: the H$_2$ formation rate and the recombination of electrons on grains.\\

\subsection{$N$(H$_3^+$) vs $\zeta$ - the simple analytic expression}

The chemical network\footnote{All reaction rates used in this paper come from KIDA database http://kida.obs.u-bordeaux1.fr \citep{Wakelam12} and the 2012 edition of the UMIST database http://udfa.ajmarkwick.net.} of \hhhp~is simple. Formation of H$_3^+$ involves the ionization of H$_2$ by cosmic rays, followed by the reaction of \hhp~with \hh~:
\begin{eqnarray}
\textrm{H}_2 + \textrm{cosmic rays} \longrightarrow \textrm{H}_2^+ + \textrm{e}^{-}  &k_{\zeta} = 0.96 \times \zeta \quad \textrm{s}^{-1},\\
\textrm{H}_2^+ + \textrm{H}_2 \longrightarrow \textrm{H}_3^+ + \textrm{H} &k_1 = 2.0 \times 10^{-9} \quad \textrm{cm}^3  \textrm{s}^{-1} 
.\end{eqnarray}
In diffuse environments, the main \hhhp~destruction reaction is the recombination with electrons, 
\begin{eqnarray}
\textrm{H}_3^+ +  \textrm{e}^{-}  \longrightarrow  products &  k_e  = 6.7\times 10^{-8} (T/300)^{-0.52} ~ \textrm{cm}^3  \textrm{s}^{-1}  
\end{eqnarray}
In dense media, destruction of \hhhp~by CO has to be considered as well.\\
 
\noindent Assuming this simple chemical network, constant densities, homogeneous medium and integrating over the whole cloud of depth $L$, the steady state \hhhp~column density $N(\textrm{H}_3^+)$ is
\begin{equation}
\label{Eq:Indriolo}
N(\hhhpm) = 0.96 \, \frac{\zeta \, L}{k_e} \frac{f}{2 \, x_e}
\end{equation}
with $f$ the molecular fraction defined as $2N(\hhm)/N_\textrm{H}$, where $N_\textrm{H}$ =  $N(\hm) + 2 \times N(\hhm)$ and 
with $x_e$~=~$n$(e$^{-}$) / $n_\textrm{H}$ the electronic fraction.\\
If the molecular fraction,  electronic fraction, and  gas temperature are known or can be guessed, one may deduce $\zeta \times L$ from the measure of $N(\hhhpm)$. In local diffuse and translucent clouds, when available, the molecular fraction is derived from observations in absorption of H and H$_2$ in the far UV, and the gas temperature is deduced from the ratio of the two first H$_2$ rotational levels J=0 and J=1 \citep{Rachford02}. The electronic fraction is usually assumed to be given by $n(\textrm{C}^+)/n_{\textrm{H}}$. This expression has been extensively used by \cite{McCall02, McCall03, Indriolo07, Indriolo10, Indriolo12} to deduce $\zeta$ in local diffuse clouds. The conclusion of these studies is $\zeta \sim 1.7 ~\times 10^{-16} - 1.1 ~\times 10^{-15}$ s$^{-1}$ \citep{Indriolo12} in diffuse clouds, a significantly larger value than previously deduced from OH and HD observations. On the other hand, a detailed simulation with the Meudon PDR code of the diffuse line of sight toward $\zeta$ Persei, where OH, HD, and {\hhhp} have been detected concomitantly, leads to a value of the cosmic-ray ionization rate of 2.5~$\times$ 10$^{-16}$ s$^{-1}$ \citep{Pap2004}, whereas the simple analytic derivation restricted to the analysis of \hhhp~concluded to a value of $\sim 1.2 \times 10^{-15}$ s$^{-1}$ \citep{McCall03}. 
The value obtained by \cite{Pap2004} relies on the modeling of 18 species detected toward $\zeta$ Persei with the Meudon PDR code \citep{Pap2006} and the Paris-Durham shock code \citep{Flower98}. One major constraint is $N(\textrm{OH}),$ which is also proportional to $\zeta$, so that the high value of $\zeta$ derived by \cite{McCall03} leads to a  column density of OH that is too large compared to observations \citep{felenbok96, roueff96}.\\

Equation \ref{Eq:Indriolo} has also been used to infer $\zeta \times L$ toward various sources located in the CMZ \citep{Oka05, Goto08, Goto11}.  Very restricted molecular information is available for these lines of sight, however; the molecular fraction is unknown and the electronic fraction is still assumed to be $n(\textrm{C}^+)/n_{\textrm{H}}$. Assuming $f$~=~1, an order of magnitude of $\zeta \sim$ (2-7) $\times$ 10$^{-15}$ s$^{-1}$ is estimated by  \cite{Oka05}.

\subsection{$N$(H$_3^+$) vs $\zeta$ relationship  - numerical models}
\label{Sec:H3pZet}

In this section, we test Eq. \ref{Eq:Indriolo} with the Meudon PDR code \citep{Pap2006} 
for diffuse cloud conditions. We consider different diffuse clouds defined by their total visual extinctions, $A_\textrm{V}^{max}$ = 1, proton densities (three densities are considered, $n_\textrm{H}$ = 50, 100, and 1000 \cmt), and illuminated on both sides by the isotropic ISRF expressed in Mathis units \citep{Mathis83}. The gas temperature is computed at each position in the cloud taking detailed cooling and heating mechanisms into account  (photoelectric effect, cosmic-ray heating, exothermic reactions, ...), as described in \cite{Pap2006} and \cite{Manuel08}. Our chemical network includes 165 species linked by 2850 chemical reactions. The code computes stationary state chemical abundances at each position using photodestruction rates determined by the radiative transfer and  gas temperature computed by the thermal balance. Then, column densities are computed. To test Eq. \ref{Eq:Indriolo} on a wide range of cosmic-ray ionization rate values, we ran models with $\zeta$ from 10$^{-17}$ to 10$^{-12}$ s$^{-1}$. The input parameters are summarized in Tables \ref{Tab:InputParam} and \ref{Tab:ElemAbund}. We also introduce the metallicity Z as a multiplicative factor of the elemental abundances of heavy elements (C, N, O, S, F, ...) and of grain abundances. In this section, Z is fixed to 1, whereas in Sect. \ref{Sec:CMZ}, which is dedicated to CMZ conditions, Z is set to 3.\\
\begin{table}[h!]
\center
\caption{Input parameters used in the Meudon PDR code common to all models discussed in Sect. \ref{Sec:H3pZet}.}
\label{Tab:InputParam}
\begin{tabular}{lll}
\hline
\hline
Parameter                & Value                      &  Unit / Ref   \\
\hline
Geometry                 & plane-parallel             &                       \\
Size                     & $A_{\textrm{V}}^{max}$=1   &                       \\
Equation of state        & isochoric                  &                       \\
Density                  & 50, 100, 1000              & cm$^{-3}$             \\
ISRF scaling factor, $G_0$  & 1 (2 sides)    &  $^{(1)}$             \\
Cosmic-ray ionization rate, $\zeta$       & variable  & s$^{-1}$              \\
Metallicity, Z           & 1                          &                       \\
Dust extinction curve    & mean Galactic              & $^{(2)}$              \\
Mass grain / mass gas    & 0.01 $\times$ Z            &                       \\ 
Mass PAH / mass grain    & $4.6 \times 10^{-2}$       & $^{(3)}$              \\
$N_\textrm{H}$ / E(B-V)  & $5.8\times 10^{21}$ / Z    & cm$^{-2}$ mag$^{-1}$  \\
R$_\textrm{V}$           & 3.1                        & $^{(4)}$              \\
Grain size distribution  & $\propto$ $r^{-3.5}$       & $^{(5)}$              \\
Minimum grain radius     & $1\times 10^{-7}$          & cm                    \\
Maximum grain radius     & $3\times 10^{-5}$          & cm                    \\
\hline
\end{tabular}
\tablebib{(1) \cite{Mathis83},  (2) \cite{Fitzpatrick86}, (3) \cite{Draine07}, (4) \cite{Seaton79}, (5) \cite{Mathis77}}
\end{table}

\begin{figure}[h]
\includegraphics[width=9.2cm]{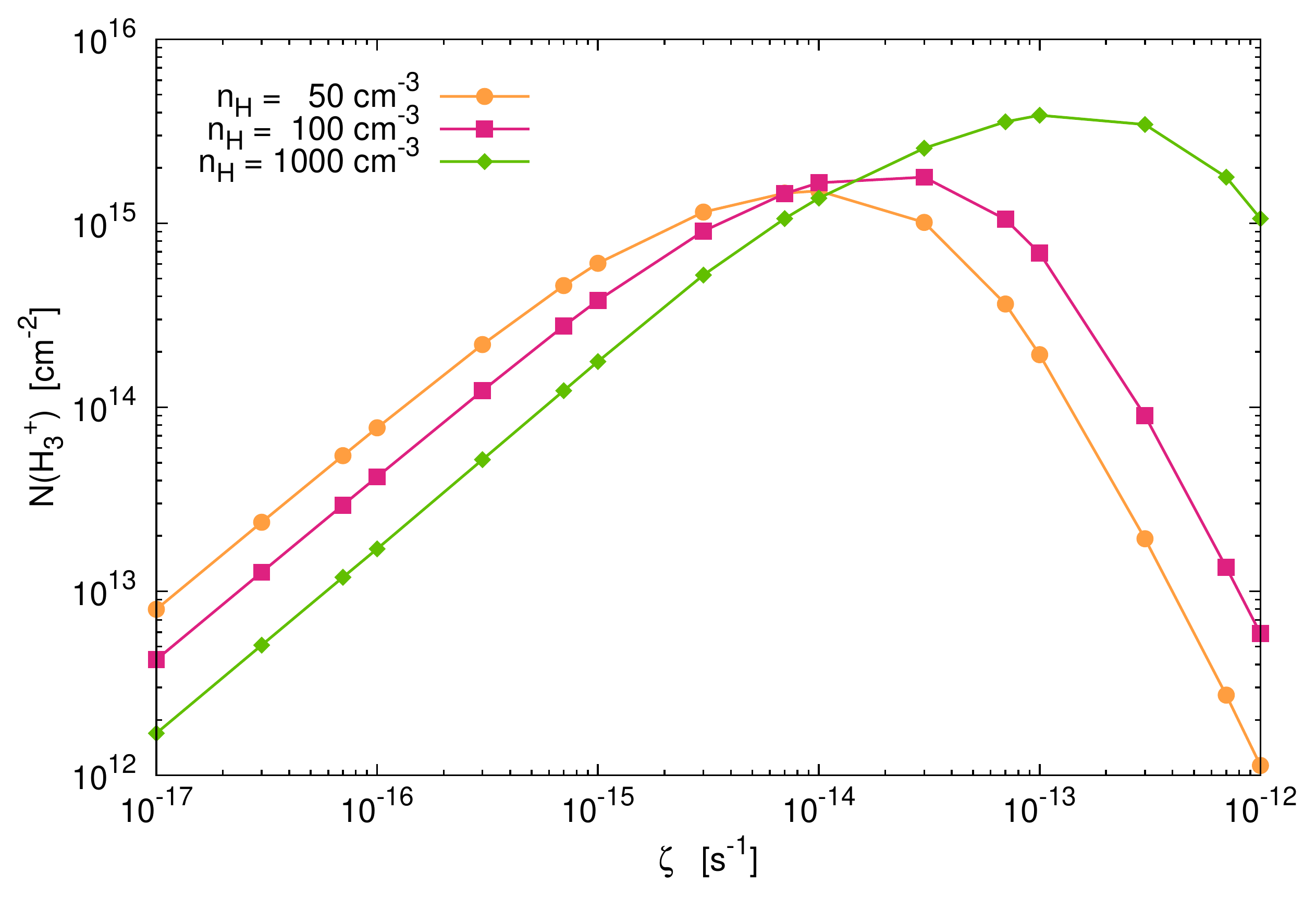}
\caption{Column density of \hhhp~as a function of $\zeta$. Each point corresponds to a PDR model as defined in Tables \ref{Tab:InputParam}
 and \ref{Tab:ElemAbund}. }
\label{Fig:NH3p_zeta}
\end{figure}

\begin{table}[h!]
\center
\caption{Elemental abundances. Z is the metallicity.}
\label{Tab:ElemAbund}
\centering
\begin{tabular}{lll}
\hline
\hline
Element  &  Elemental abundance           & Reference \\
\hline
He    & 0.1                               &   \\
C     & $1.32\times 10^{-4}$ $\times$ Z   & 1 \\
O     & $3.19\times 10^{-4}$ $\times$ Z   & 2 \\                     
S     & $1.86\times 10^{-5}$ $\times$ Z   & 1 \\
N     & $7.50\times 10^{-5}$ $\times$ Z   & 3 \\
F     & $1.8\times 10^{-8}$ $\times$ Z    & 4 \\
\hline
\end{tabular}
\tablebib{(1) \cite{Savage96},(2) \cite{Meyer98}, (3) \cite{Meyer97}, (4) \cite{Snow07}}
\end{table}
$N(\textrm{H}_3^+)$ as a function of $\zeta$ is presented in Fig. \ref{Fig:NH3p_zeta} where each point corresponds to a model. 
\hhhp~column density increases with $\zeta$ only up to a maximum value of the cosmic-ray ionization rate, $\zeta^{max}$. For moderate $\zeta$, the relationship is linear, in agreement with Eq. \ref{Eq:Indriolo}. Using the computed $f$, $x_e$, and $T$, we checked that Eq. \ref{Eq:Indriolo} gives an extremely good agreement with the numerical model in the linear increasing part. However, when Eq. \ref{Eq:Indriolo} is used to estimate $\zeta$ with observations, the molecular fraction, electronic fraction, and gas temperature may have to be guessed, contrary to our numerical check where they are consistently computed. The maximum of N(H$_3^+$), reached at $\zeta = \zeta^{max}$, depends on the gas density. In these models, the order of magnitude of the maximum of $N(\textrm{H}_3^+)$ is 10$^{15}$ cm$^{-2}$. This value depends on the size of the cloud, here $A_{\textrm{V}}^{max}$ = 1, and on the gas temperature (see below). For cosmic-rays ionization rates higher than (or even close to) $\zeta^{max}$, Eq. \ref{Eq:Indriolo} is obviously incorrect and cannot be used to deduce $\zeta$ from $N(\textrm{H}_3^+)$.
This decrease of \hhhp~column density with the cosmic-ray ionization rate can be understood from the behavior of the molecular and electronic fractions with $\zeta$.\\ 

\begin{figure}[h!]
\includegraphics[width=9.2cm]{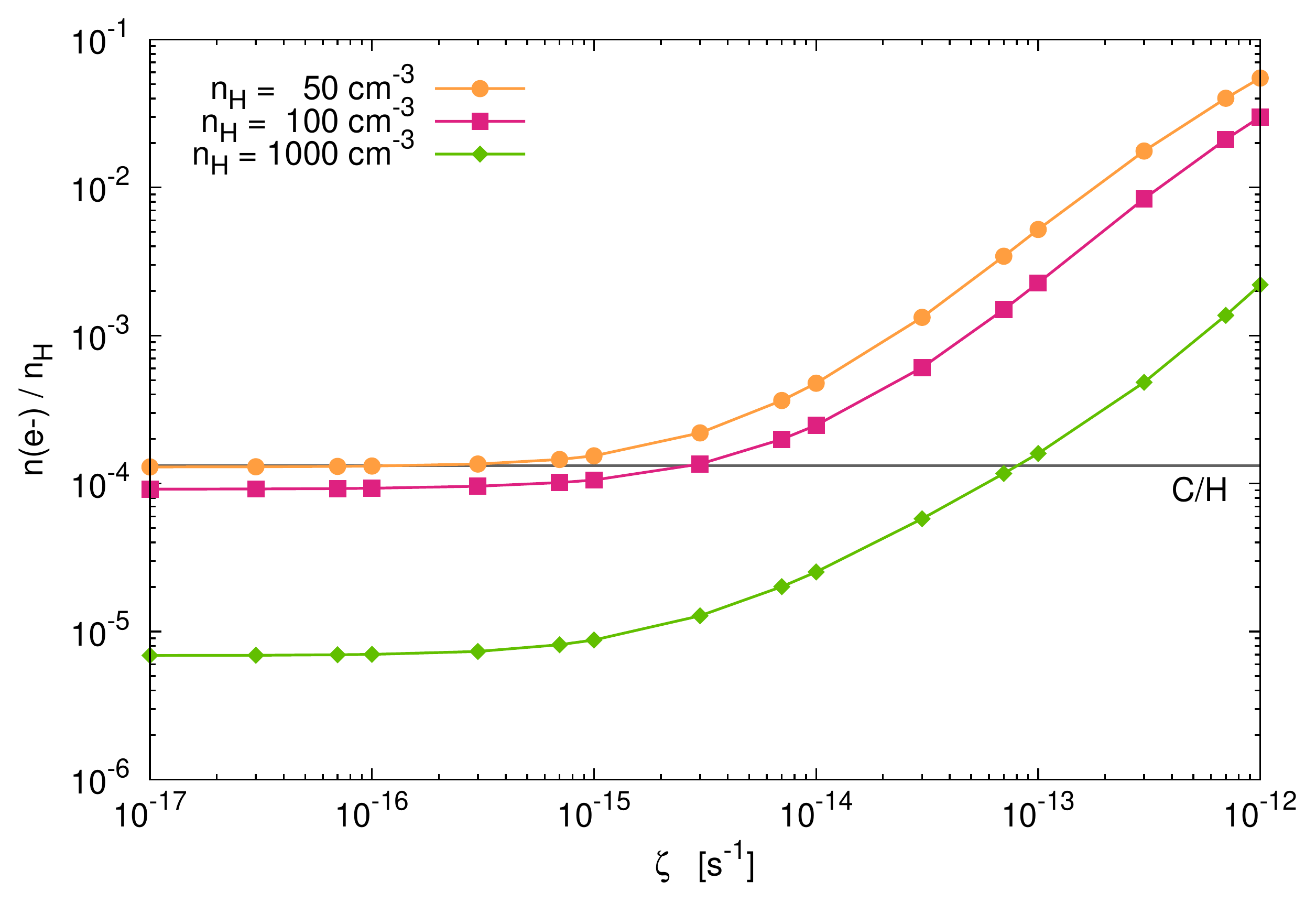}
\caption{Computed electronic fraction, $x_e$, at \av~=~0.5 as a function of $\zeta$. Each point corresponds to a PDR model as defined in Tables \ref{Tab:InputParam} and \ref{Tab:ElemAbund}. The horizontal line represents the elemental abundance of carbon relative to H.}
\label{Fig:Abelectrons}
\end{figure}

Figure \ref{Fig:Abelectrons} shows the computed electronic fraction, $x_e$ = $n$(e$^-$)/\nh, in the middle of the cloud ($A_\textrm{V}$ = 0.5) as a function of $\zeta$. The elemental abundance of C, C/H, is a good proxy for $x_e$ for low values of $\zeta$ and for low density clouds (50 and 100 cm$^{-3}$ in our examples). At high cosmic-ray fluxes, $x_e$ is significantly higher than the C/H ratio because a significant amount of electrons is produced by the cosmic-ray ionization of H and H$_2$. At high densities and low cosmic-ray fluxes, $x_e$ is lower than the C/H ratio because of efficient neutralization of electrons on grains and PAHs. This point has been discussed by \cite{Liszt03, Liszt06}.\\

\begin{figure}[h!]
\includegraphics[width=9.2cm]{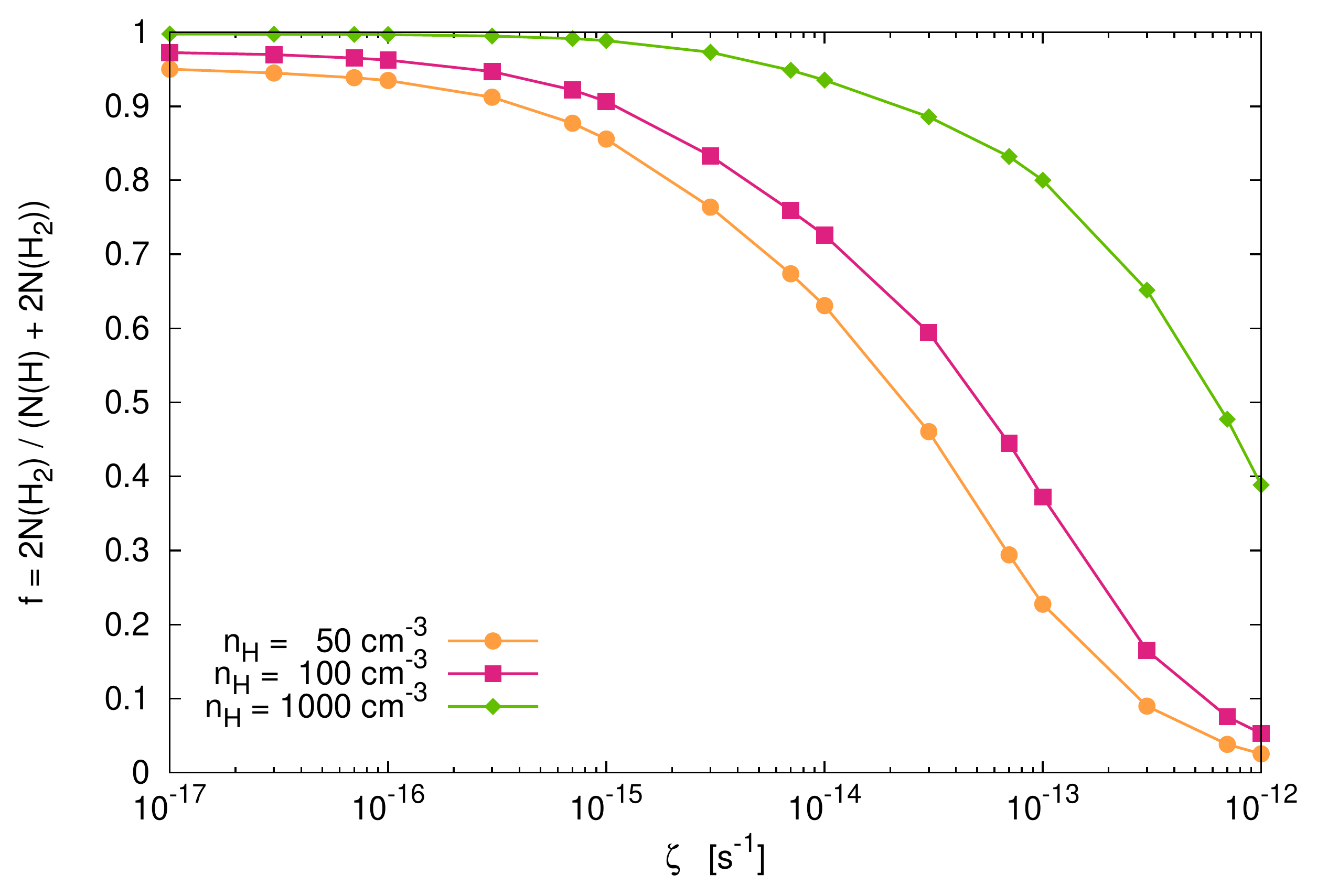}
\caption{Molecular fraction $f$ as a function of $\zeta$. Each point corresponds to a PDR model as defined in Tables \ref{Tab:InputParam}
 and \ref{Tab:ElemAbund}. }
\label{Fig:MolFrac}
\end{figure}

Figure \ref{Fig:MolFrac} illustrates the decrease of the molecular fraction with $\zeta$. The lower the density of the gas, the lower the $\zeta$ at which this decrease happens. At high flux of cosmic rays, molecular hydrogen is efficiently ionized by cosmic rays and forms H$_2^+$, which  quickly reacts with electrons to dissociate in hydrogen atoms or reacts with H and H$_2$ to produce H$_2$ and H$_3^+$. Molecular fractions computed by the PDR code reach values up to 0.9 - 1, whereas FUSE observations of local diffuse clouds show smaller molecular fractions, $f\sim 0.6$ \citep{Rachford02}. The molecular fractions presented here correspond to single-cloud models with a total visual extinction of 1. It is possible that several small components are present on FUSE lines of sight, resulting in a smaller total molecular fraction. This is consistent with a picture of fragmented diffuse interstellar gas  \citep{Godard14}.\\

The dependence of $N(\textrm{H}_3^+)$ with $\zeta$ is then straightforward to understand. When the cosmic-ray ionization rate is large, $N(\textrm{H}_3^+)$ decreases with $\zeta$, first, because less H$_2$ is available to form H$_3^+$ and, second, because H$_3^+$ recombines efficiently because of a large abundance of electrons produced by cosmic-ray ionization of H and H$_2$. A more general relationship between $N(\textrm{H}_3^+)$ and $\zeta$ than Eq. \ref{Eq:Indriolo} can be obtained taking into account, in the chemical balance, H$_2^+$ reactions that are important at large $\zeta$:  
\begin{equation*}
\textrm{H}_2^+ +  \textrm{e}^- \longrightarrow  \textrm{H} + \textrm{H} \quad \quad k_e(\textrm{H}_2^+) = 2.53\times 10^{-7}~ (T/300)^{-0.5}~ \textrm{cm}^{3} \textrm{s}^{-1}
\end{equation*}
\begin{equation*}
\textrm{H}_2^+ +  \textrm{H} \longrightarrow \textrm{H}_2 + \textrm{H}^+   \quad k_2 = 6.4\times 10^{-10}~ \textrm{cm}^{3} \textrm{s}^{-1}.
\end{equation*}

Then, the column density of \hhhp~in diffuse gas is written 
\begin{eqnarray}
\label{Eq:nH3p}
N(\hhhpm) = 0.96 \times \frac{\zeta L}{k_e} \,\, \frac{f}{2 x_e} \left[ 1 + \frac{2 \,\, k_e(\textrm{H}_2^+) \,\, x_e}{k_1 \,\, f} + \frac{2k_2}{k_1} \left( \frac{1}{f} - 1\right)\right]^{-1}
\end{eqnarray}
A similar equation is given by \cite{Indriolo12} where destruction of \hhhp~by CO is included. Equation \ref{Eq:nH3p} shows that, as long as the expression in brackets is close to 1 (low $\zeta$), we recover the linear dependence between the column density of \hhhp~and $\zeta \times L$. However, when dissociative recombination of \hhp~and charge transfer with H compete with destruction of \hhp~by H$_2$, Eq. \ref{Eq:nH3p} shows that the column density of \hhhp~is reduced compared to the prediction of Eq. \ref{Eq:Indriolo}.\\

Cosmic-ray ionization rates can thus only be deduced with the classical analytical expression, Eq. \ref{Eq:Indriolo}, for moderate $\zeta$, as in standard diffuse lines of sight. For large $\zeta$, this relationship fails. A more general expression, appropriate for low and high cosmic-ray fluxes in diffuse environments, is Eq. \ref{Eq:nH3p}. In both cases, the difficulty to use such analytical expressions is to estimate properly $f$ and $x_e$. 

\subsection{Molecular fraction and H$_2$ formation processes}
\label{Sec:H2form}

Because the molecular fraction is a key parameter in the relationship between the cosmic-ray ionization rate and $N(\textrm{H}_3^+)$, the transition from atomic to molecular gas must be computed properly. The analytic theory of the H-H$_2$ transition has been described by \cite{Sternberg14}. In diffuse gas, H$_2$ is destroyed by UV photons via absorption transitions in the Lyman and Werner bands followed by de-excitation in the continuum of the ground electronic state. In the models presented here, this process is computed taking  continuum absorption by dust and carbon atoms of the UV radiation field and self-shielding of H$_2$ lines into account\citep{Pap2006}. The formation of H$_2$ takes place on grains. In numerical models, this can be simulated in very different ways, from simple analytic formulae to very detailed modeling. In most astrochemical models, a mean formation rate of $3\times10^{-17} \sqrt(T/100)$ cm$^{3}$ s$^{-1}$ is used. This rate was deduced from Copernicus and FUSE observations of diffuse clouds in the local neighborhood \citep{Jura74, Gry02}. Several studies based on ISO and Spitzer observations showed that this rate is too low to account for H$_2$ emission lines in PDRs where the gas and the grains are warm \citep{Habart04, Habart11}. One may wonder if the mean formation rate determined in local diffuse lines of sight also applies  to the CMZ, where the gas is warm too. Indeed, this mean value hides complex microscopic mechanisms: H adsorption in chemisorption and physisorption sites, migration at the surface of grains and H$_2$ formation. The mean value also hides the properties of grains and PAHs, such as their composition and temperature. In the Meudon PDR code, several formalisms are implemented to simulate H$_2$ formation on grains: analytic expression \citep{Pap2006}, moment equation formalism \citep{Pap2009}, Langmuir-Hinshelwood (LH) and Eley-Rideal (ER) mechanisms \citep{LeBourlot12}, and a stochastic approach that considers the impact of grain temperature fluctuations on H$_2$ formation rate  \citep{Bron14}.\\

In all models presented here, the H$_2$ formation rate is computed using the formalism described in \cite{LeBourlot12}, i.e., we take  into account adsorption in physisorption and chemisorption sites with H$_2$ formation via LH and ER mechanisms\footnote{We have not used the most sophisticated H$_2$ formation model at our disposal \citep{Bron14}, which  takes  the effect of grain temperature fluctuations on H$_2$ formation into account because it is too  CPU time consuming. Nevertheless, we checked that grain temperature fluctuations do not have a strong effect on our conclusions.}. We upgraded the grain model assuming the grain size distribution contains a log-normal PAH component \citep{Compiegne11} plus a MRN power law \citep{Mathis77} for amorphous carbons and silicates with minimum and maximum radius equal to $10^{-7}$ and 3$\times10^{-5}$ cm. H$_2$ formation rate on PAHs is not well known, but several laboratory experiments show that the process is efficient \citep{Boshman12, Mennella12}. Here, we assume that ER mechanism on PAHs is as efficient as on grains. As emphasized in \cite{LeBourlot12}, the efficiency of the LH mechanism depends on grain temperatures whereas the efficiency of the ER mechanism depends on the gas temperature. Indeed, only H atoms with sufficient kinetic energy can reach chemisorption sites. In our models, we assume a threshold of 300 K to reach these sites. This low value accounts for the likely presence of defects on grain surfaces.\\ 

\begin{figure}[h!]
\includegraphics[width=9.2cm]{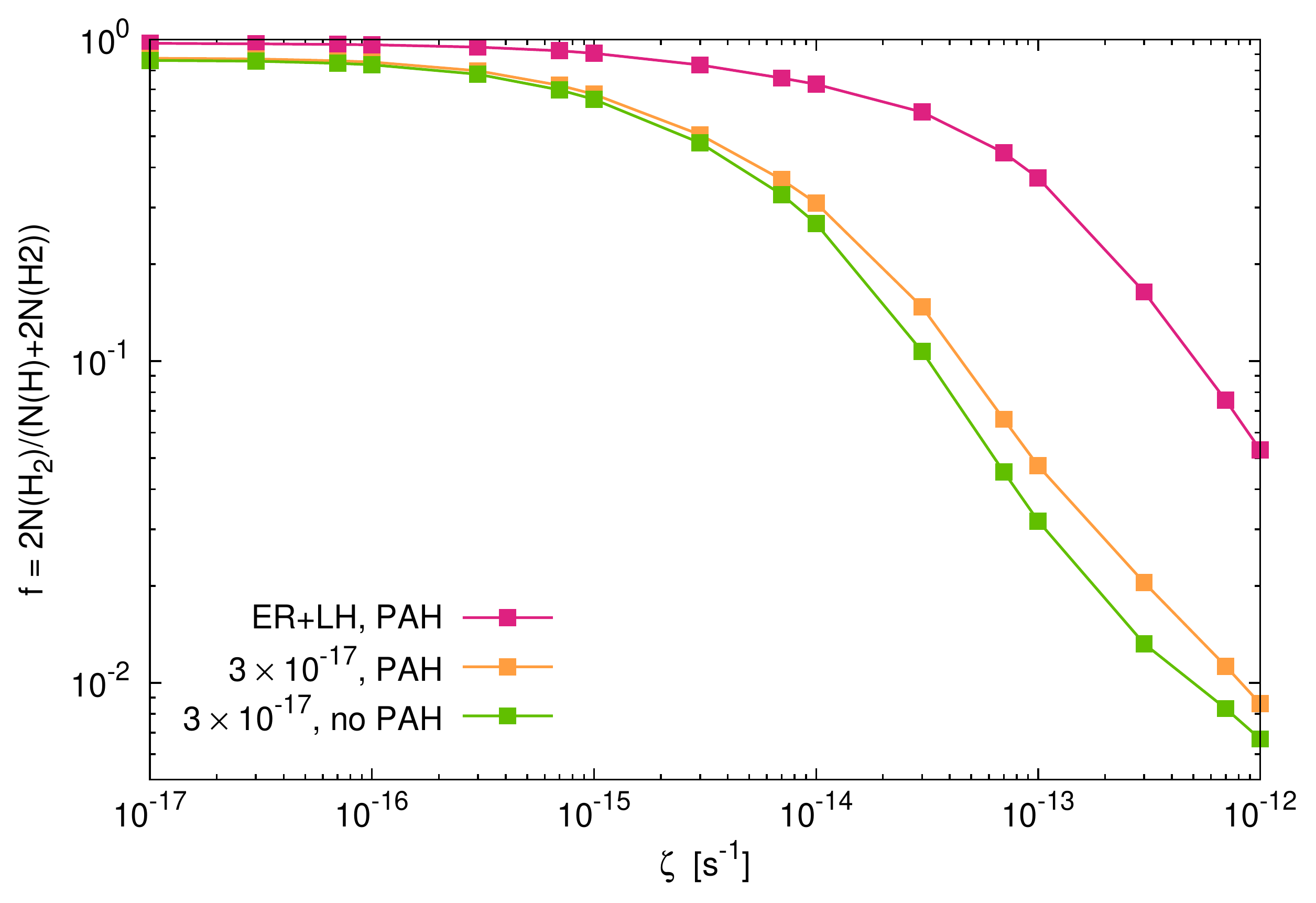}
\caption{Molecular fraction as a function of $\zeta$ for different grain distributions and H$_2$ formation mechanisms. Each point corresponds to a PDR model as defined in Tables \ref{Tab:InputParam} and \ref{Tab:ElemAbund} with $n_\textrm{H}$ = 100 cm$^{-3}$.}
\label{Fig:EffectERPAH}
\end{figure}

To study the sensitivity of the computed molecular fraction to the H$_2$ formation model, we ran the same models as in the previous section with two prescriptions for the H$_2$ formation rate on grains: the crude approximation, $3\times10^{-17} \sqrt(T/100)$ cm$^{3}$ s$^{-1}$, and the more physical model taking  ER and LH mechanisms into account. Grain and PAH properties in the CMZ are poorly known. Since they affect the photoelectric heating and ionization degree through ions and electrons recombination and H$_2$ formation, we also ran models with and without the PAH component. Computed $f$ as a function of $\zeta$ are presented for the $n_\textrm{H}$ = 100 cm$^{-3}$ models in Fig.~\ref{Fig:EffectERPAH}. In models with fixed H$_2$ formation rate, the molecular fraction does not depend on the grain surface. There is only a slight $f$ increase at high $\zeta$ (between PAH and no PAH models) because of an increase of the gas temperature and variations of the available grain surface for charge recombinations. On the the other hand, the consideration of ER mechanism leads to significantly higher molecular fractions at high $\zeta$ than the $3\times10^{-17} \sqrt(T/100)$ approximation. 

\begin{figure}[h!]
\includegraphics[width=9.2cm]{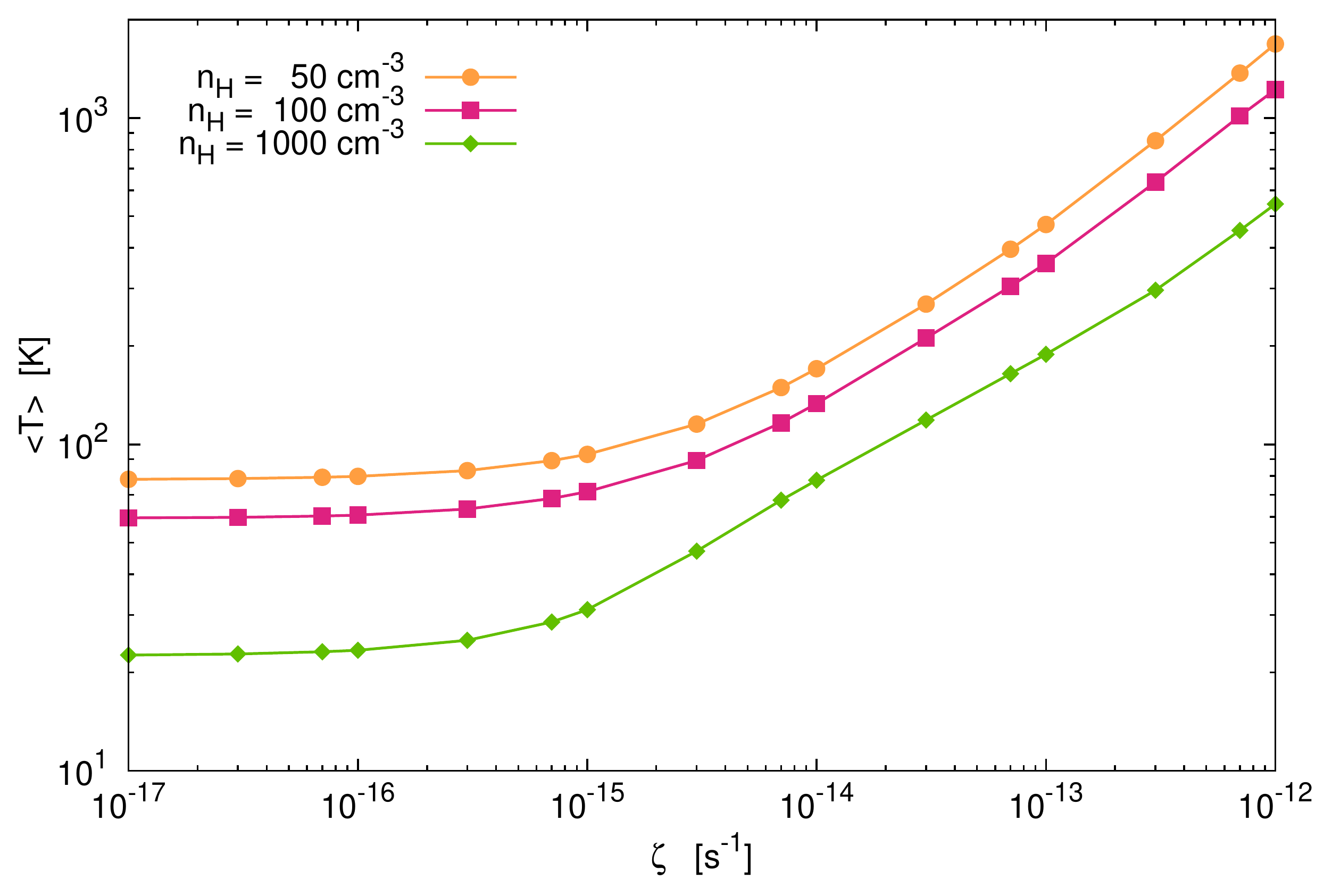}
\caption{Mean gas temperature for all models presented in Sect. \ref{Sec:H3pZet} Each point corresponds to a PDR model as defined in Tables \ref{Tab:InputParam} and \ref{Tab:ElemAbund}.}
\label{Fig:GasTemp}
\end{figure}

Figure \ref{Fig:GasTemp} shows the gas temperature averaged over positions, <$T$>, for all models presented in Sect. \ref{Sec:H3pZet}. For $\zeta < 10^{-15}$ s$^{-1}$ and diffuse conditions (n$_\textrm{H} \le 100$ \cmt), <$T$> is $\sim$ 60 K, in agreement with FUSE observations \citep{Rachford02}. Increasing $\zeta$ from 10$^{-17}$ to 10$^{-12}$ s$^{-1}$ increases the gas temperature from 60 to 1200 K in models with $n_\textrm{H}$ = 100 cm$^{-3}$. In models with a large value of $\zeta$, cosmic rays heat the gas to several hundred Kelvin, firstly, by direct ionization of H and H$_2$, followed by thermalization of electrons with the gas; and, secondly, because electrons recombine in exothermic dissociative recombination reactions (mostly the H$_3^+$ + e$^-$ reactions with $\Delta E$ = 4.7 eV in the three body dissociation). In our models, we assume that each cosmic-ray ionization contributes to the heating of the gas by 4 eV. Usually, formation of H$_2$ via chemisorbed H atoms is efficient only in photodominated regions, where the temperature of the gas reaches several hundred Kelvin. In typical diffuse clouds, where the gas is at $\simeq$~60 K \citep{Rachford02}, this process is unlikely. The situation may be different for diffuse gas in the CMZ for which \cite{Oka04} suggested that the temperature is several hundred Kelvin, enough to provide the kinetic energy for H atoms to reach chemisorption sites and increase the H$_2$ formation rate by a significative amount compared to local diffuse clouds.\\ 

\subsection{Electronic fraction and e$^-$ recombinations on grains}

The second important parameter in deriving  $\zeta$ from \hhhp~observations is the electronic fraction (cf. Eqs. \ref{Eq:Indriolo} and \ref{Eq:nH3p}). We showed that, at high  cosmic-ray flux, electrons produced by ionization of H and H$_2$ cannot be neglected compared to those produced by UV ionization of carbon and grains.\\

\begin{figure}[h!]
\includegraphics[width=9.2cm]{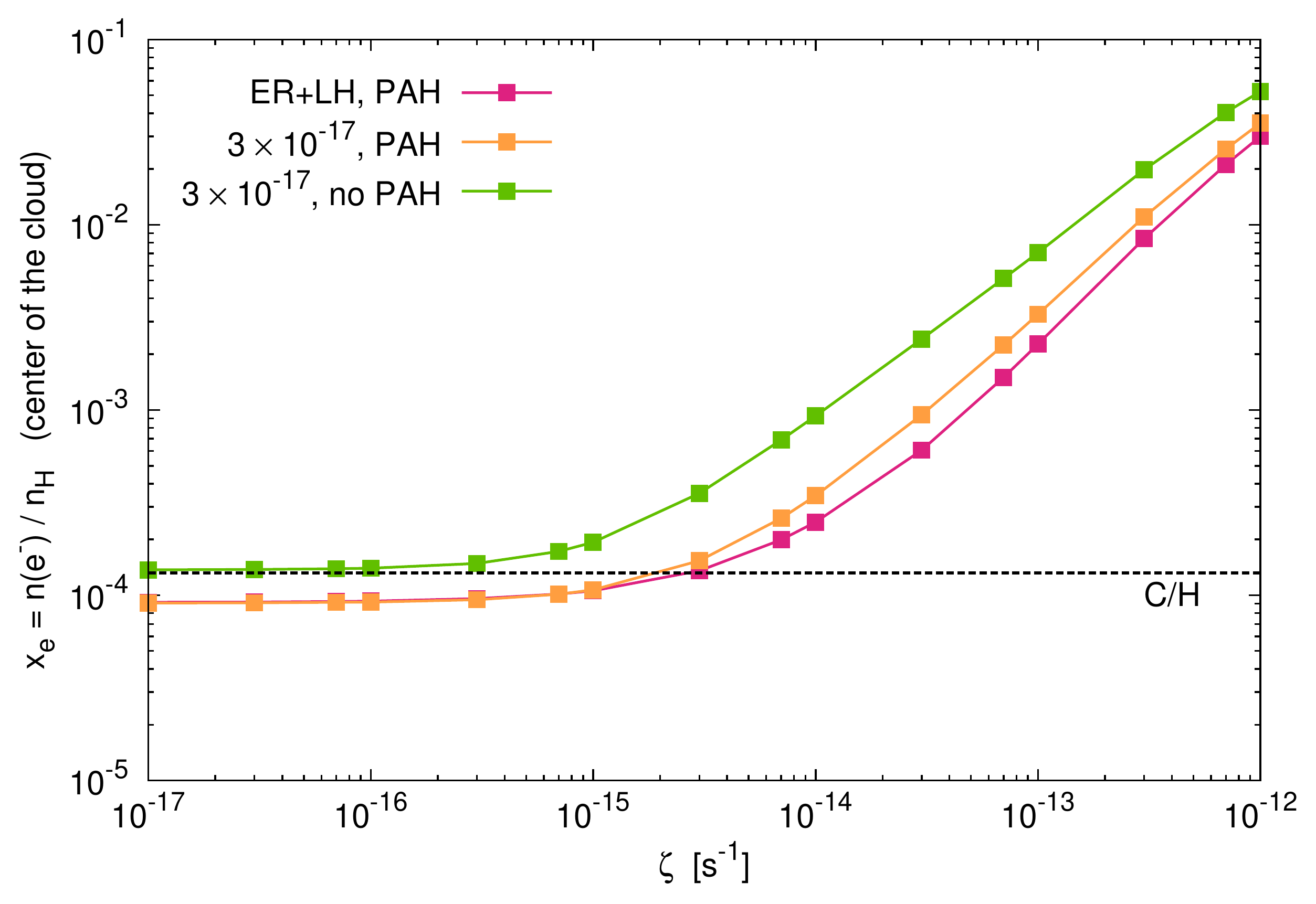}
\caption{Electronic fraction as a function of $\zeta$ for different grain distributions and H$_2$ formation mechanisms. Each point corresponds to a PDR model as defined in Tables \ref{Tab:InputParam} and \ref{Tab:ElemAbund} with $n_\textrm{H}$ = 100 cm$^{-3}$.}
\label{Fig:EffectERPAH_ElFrac}
\end{figure}

Figure \ref{Fig:EffectERPAH_ElFrac} presents $x_e$ as a function of $\zeta$ for the previous \nh~=~100~\cmt~models. The H$_2$ formation formalism has only  a small impact on $x_e$. On the contrary, grain surface has a strong impact on $x_e$. The higher the available grain surface to recombine electrons, the lower the electronic fraction. We do not know  the grain properties in the CMZ. In our comparison of models to CMZ observations, we  assume a standard grain size distribution plus a PAH component, as described in Sect. \ref{Sec:H3pZet} and \ref{Sec:H2form} and we  scale the grain mass with the metallicity in the CMZ.\\

\subsection{Effect of the gas temperature}

Warm gas contributes to increase the abundance of \hhhp~\citep{Pap2004}. First, as described above, it increases H$_2$ formation rate by ER mechanism on chemisorption sites. The available H$_2$ may lead to the formation of H$_3^+$. Second, the dissociative recombination rate of H$_3^+$ decreases with $T$. Third, at $T$ above 100 K, the charge exchange reaction between O and H$^+$ becomes efficient, i.e.,
\begin{equation*}
\textrm{H}^+ ~ + ~ \textrm{O} \longrightarrow \textrm{O}^+~ +~ \textrm{H} \quad k = 7.31\times 10^{-10} (T/300)^{0.23} e^{-226/T},
\end{equation*}
with $k$ in cm$^{3}$ s$^{-1}$. This opens the chemical network of oxygen hydride ions OH$^+$, H$_2$O$^+$, and H$_3$O$^+$ formed through successive reactions with H$_2$. An increase of the densities of these hydrides provides additional channels for electrons consumption. Hence, at large temperatures, electrons have several ways to recombine efficiently in competition with H$_3^+$ recombination. Moreover, it is then not surprising to find similar velocity profiles in H$_3^+$ and H$_2$O$^+$ spectra obtained in two nearby CMZ sources, the $\iota$ star in the Galactic center and Sgr B2 \citep{Schilke13}, as emphasized by \cite{Oka15}. We  specifically discuss the link between H$_3^+$ and oxygen hydride ions in Sect. \ref{Sec:OHp}. As a result of these temperature dependence effects on H$_3^+$, it is also less surprising to observe large H$_3^+$ column densities in the warm diffuse gas of the CMZ.  

\subsection{Conclusion on $N(\textrm{H}_3^+)$ and $\zeta$}

\begin{figure}[h!]
\includegraphics[width=9.2cm]{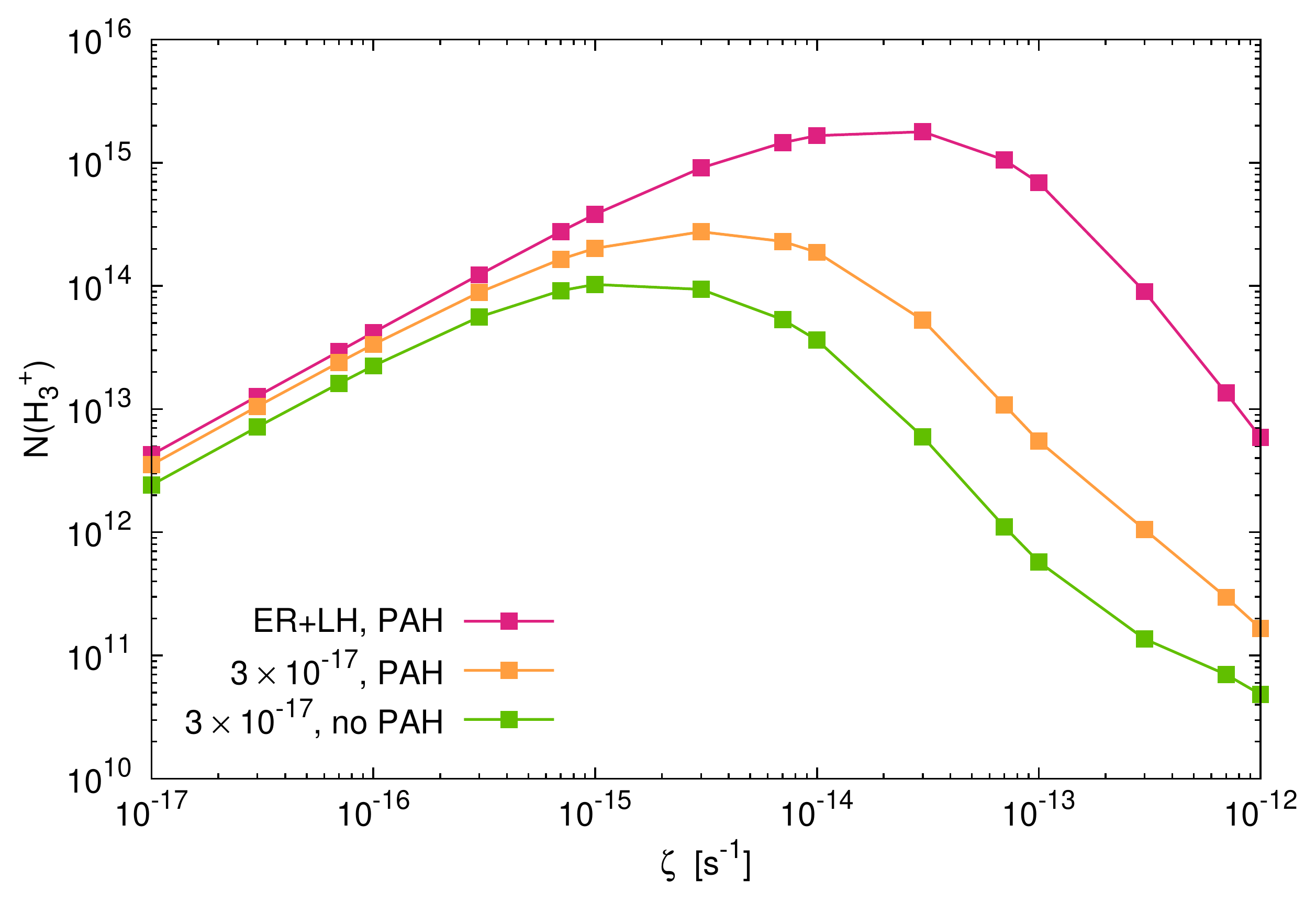}
\caption{$N(\textrm{H}_3^+)$ as a function of $\zeta$ for the two H$_2$ formation models ($3\times10^{-17} \sqrt(T/100)$ cm$^{3}$ s$^{-1}$ and Langmuir-Hinshelwood plus Eley-Rideal mechanisms) and two grain distributions: MRN-like distribution and MRN-like + PAH distribution. For all models \nh~=~100~\cmt.}
\label{Fig:EffectERPAH_NH3p}
\end{figure}

To summarize the effect of H$_2$ formation prescription and grain / PAHs properties, Fig. \ref{Fig:EffectERPAH_NH3p} presents the dependence of $N(\textrm{H}_3^+)$ on $\zeta$ for the \nh~=~100~\cmt~models. The position of the maximum of the $N(\textrm{H}_3^+)-\zeta$ relationship depends strongly on the H$_2$ formation prescription and on grain / PAHs characteristics. A too simplistic recipe, such as the $3\times10^{-17} \sqrt(T/100)$ value, leads to erroneous conclusions when trying to deduce $\zeta$ from H$_3^+$ observations for high values of $\zeta$. This illustrates the crucial need, in interstellar chemical studies, to account for  detailed microphysical processes.\\

\section{H$_3^+$ excitation}
\label{Sec:Excit}

Gas density and temperature in diffuse gas can be inferred from \hhhp~observations in states (3,3), (2,2), and (1,1) \citep{Oka04}. \cite{Geballe12} stated that $N(3,3)$/$N(1,1)$ is a good thermometer for any temperature and $N(3,3)$/$N(2,2)$ is a good densimeter for high temperature; see  discussion in \cite{Oka13}. \cite{Oka04} computed maps of these ratios in the plane $n(\textrm{H}_2$)-$T$ and deduced that the gas probed by \hhhp~in the CMZ is diffuse ($n(\textrm{H}_2)\leq 70$ \cmt) and warm ($T \ge$ 300 K). Their computation takes  radiative de-excitations and collisional excitation and de-excitation with \hh\ into account. As \hhhp~collision rates were not available, \cite{Oka04} proposed an expression based on the Langevin expression with proper account of microreversibility. Recent theoretical studies have been performed on the \hhhp~-~\hh~system. First, \cite{Hugo09} computed reactive and nonreactive collision  rates for  temperatures below 50 K with a strong ergodicity and full nuclear spin scrambling hypothesis. Second, \cite{Roncero12} extended the temperature range up to 500 K and used a dynamically biased statistical model coupled to a recent global potential energy surface of the H$_5^+$ system \citep{Aguado10}. Finally, \cite{Park07} studied the ortho-para conversion in the \hhhp~+ H$_2$ reaction.\\

Apart from thermal processes, two additional excitation mechanisms of the (3,3) metastable level may have to be considered. As reported by \cite{Goto08}, John Black suggested the possibility of IR pumping of the ortho (1,0) that could decay after several transitions in the (3,3) metastable level. Another excitation mechanism could be the formation of \hhhp~in excited states as a result of the exothermic H$_2^+$~+~H$_2$ reaction.\\

In this section, we examine the relevance of various excitation mechanisms and discuss the relationship between \hhhp~excitation and the physical conditions of the gas, temperature, and density. We introduced the various  \hhhp~excitation mechanisms in the Meudon PDR code. Radiative data come from \cite{Lindsay01} and are available electronically on the website maintained by B. McCall (http://h3plus.uiuc.edu). To discuss the uncertainties arising from the uncertainties in the collision rates, we implemented  the two sets of collision rates by \cite{Oka04} and \cite{Roncero12}. Whereas \cite{Oka04} do not discriminate between ortho and para levels, \cite{Roncero12} consider the appropriate nuclear spin restrictions and provide state-to-state collision rates for the 24 first levels of H$_3^+$ with H$_2$. We implemented collisions of \hhhp~with \hh~in v~=~0, J~=~0 and v~=~0, J~=~1 since most of molecular hydrogen molecules are in these low energy states. We always use 24 levels for H$_3^+$ except in Sect. \ref{SubSec:IRPumping} where 54 levels are introduced, as required for our analysis. Collision rates with H and He are derived from those with \hh~J=0 with a scaling law involving the reduced mass. Since H$_3^+$ is coexistent with H$_2$, H$_2$ is the main collisional excitation and de-excitation partner. Collisions with electrons have been studied \citep{Faure03,Faure06, Kokoouline10} and may contribute to H$_3^+$ excitation. Because of the electronic potential interaction symmetries, collisional excitation due to electrons may  only occur between levels (J,K) with the same values of K. As the (3,3) metastable level is the lowest level with K~=~3, contribution of electron collisions may only take place via (4,3) de-excitation involving an energy difference of 494 K. In the present study, we did not include those collisions in the excitation balance.\\

\begin{figure}[h!]
\includegraphics[width=9.2cm]{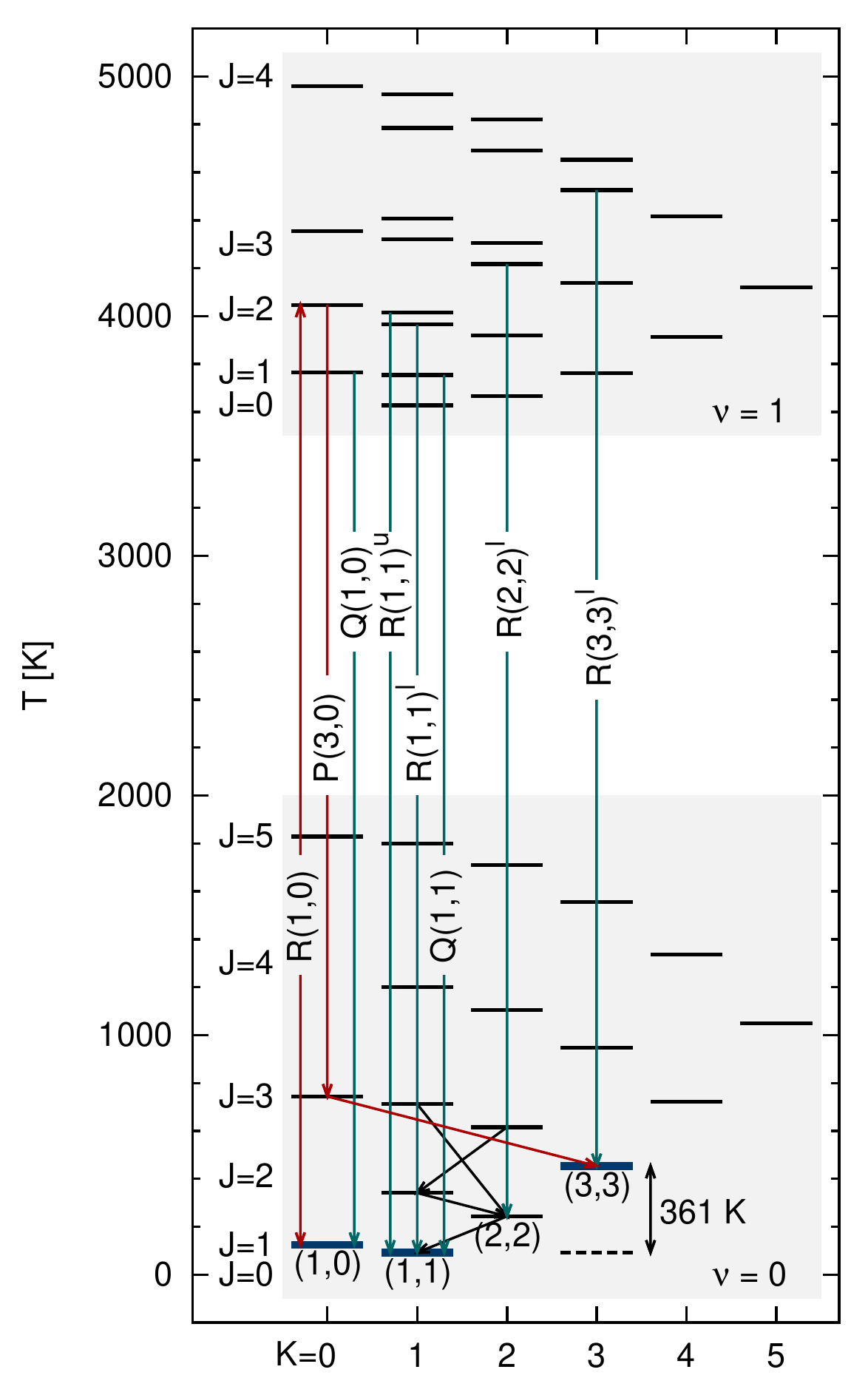}
\caption{Schema of H$_3^+$ levels with  the IR pumping mechanism discussed in Sect. \ref{SubSec:IRPumping} in red.}
\label{Fig:LevelsH3p}
\end{figure}

We recall that the Meudon PDR code computes the atomic and molecular structure of a 1D plane-parallel slab of dust and gas. By default, gas and grain temperatures are computed at each position in the cloud taking the various local heating and cooling mechanisms into account  in parallel to level excitation of several atoms and molecules responsible for the cooling of the gas. Level excitation is computed by solving the radiative transfer equation as described in \cite{Manuel08}. The PDR code considers collisional excitation and de-excitation, radiative de-excitation, radiative pumping, chemical formation, and destruction. 
It is also possible to run isothermal models with a fixed  gas temperature. This alternative is used in Sect. \ref{SubSec:Collision} to produce maps in the plane \nh~-$T$, which can be compared to \cite{Oka04}. 

\subsection{Chemical excitation}
\label{Sec:ExChim}

The exothermicity of the \hh~+~\hhp~reaction is 1.7 eV. This energy may be redistributed as kinetic energy of products and internal energy of H$_3^+$ and can contribute to the excitation of its (3,3) level. This mechanism may be efficient if the timescale associated with the formation-destruction cycle is short compared to other (de)excitation mechanisms. The chemical timescale of H$_3^+$  is inversely proportional to the product of the destructive recombination rate and H$_3^+$ density and thus dependent on the cosmic ionization rate. In the center of a stationary diffuse cloud ($n_\textrm{H}$ = 100 cm$^{-3}$, ISRF scaling factor $G_0$ = 1, $A_\textrm{V}^{max}$ = 1), we obtain a chemical timescale of about 19 yrs  for $\zeta$ = 10$^{-16}$ s$^{-1}$ and 2 yrs for $\zeta$ = 10$^{-13}$ s$^{-1}$. For comparison, the collisional de-excitation rate of the (3,3) level is found to be $\sim$ 3 yrs and $\sim$ 4 yrs, respectively. Then, chemical excitation may contribute and even dominate \hhhp~excitation at large $\zeta$.\\

Since H$_3^+$ excitation at formation is unknown, we consider three different prescriptions :
\begin{itemize}
\item[$\bullet$] {\it{Scenario A}}: we assume equipartition of energy, i.e., 33\% of the available 1.7 eV are included as internal energy (corresponding to 6510 K). This energy is far above the 24 first energy levels for which excitation rates are available \citep{Roncero12}. In addition, these excited levels decay radiatively at a rate $\sim$10$^{-3}$ s$^{-1}$ \citep{Lindsay01}, much more efficiently than through any collisional deexcitation process. We then assume that these high levels radiatively cascade toward lower levels and finally contribute  to the population of the 24 first levels proportionally to their statistical weights.
\item[$\bullet$] {\it{Scenario B}}: H$_3^+$ is formed following a Boltzmann distribution at gas temperature. This hypothesis assumes the newly formed molecule has time to thermalize with the gas. 
\item[$\bullet$]  {\it{Scenario C}}: formation energy is transferred mainly into kinetic energy  and H$_3^+$ molecules  are formed  in their para (1,1) / ortho (1,0) ground states, in the ratio of their statistical weights. 
\end{itemize}

\begin{figure}[h!]
\includegraphics[width=9.2cm]{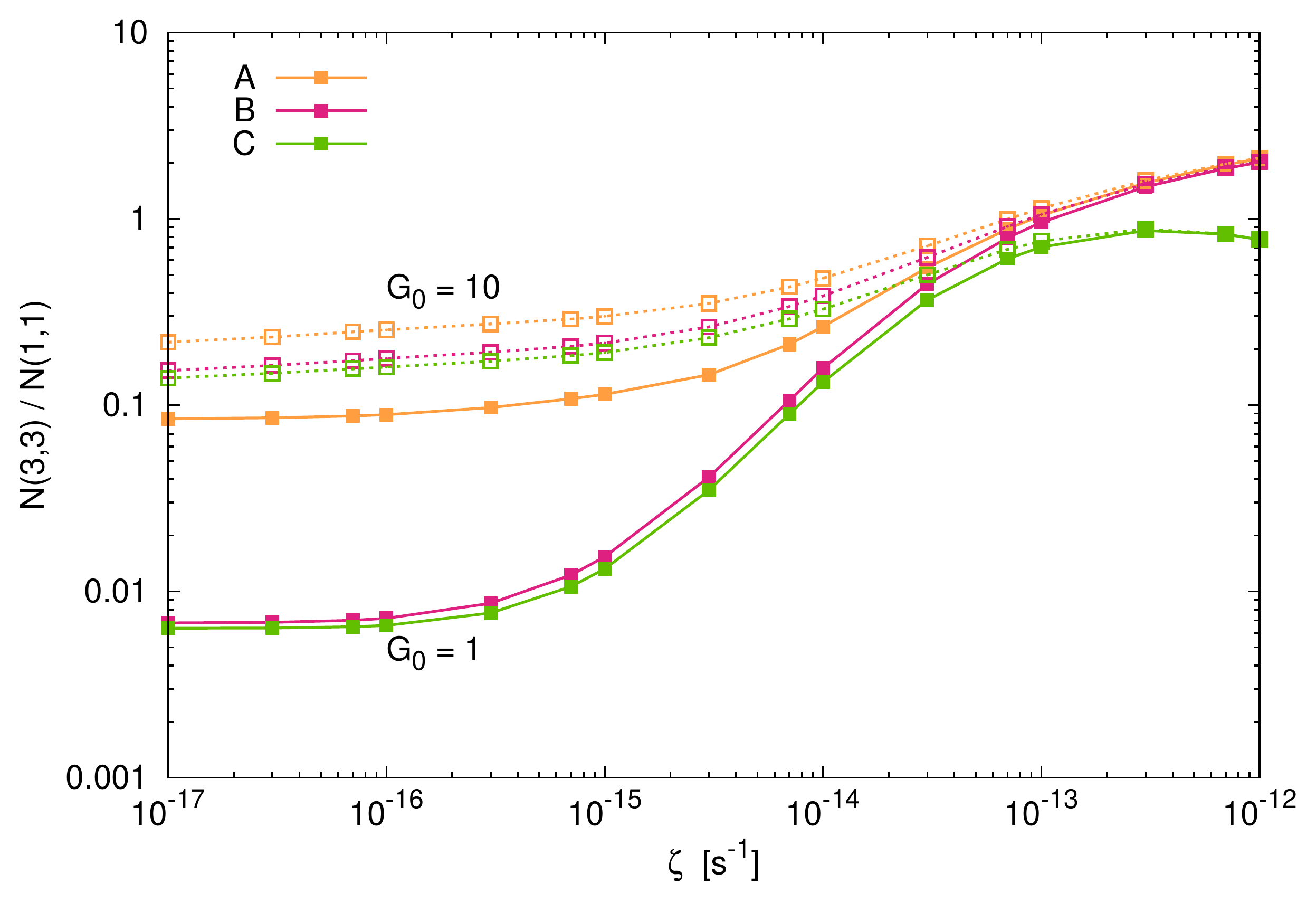}
\caption{Ratios $N(3,3)$/$N(1,1)$ as a function of $\zeta$ for three scenarii for H$_3^+$ excitation at formation. A: 33\% of the exothermicity is used as internal energy, B: H$_3^+$ is formed following a Boltzmann distribution at gas temperature, C: H$_3^+$ is formed in its two levels (1,1) and (1,0) in a ratio corresponding to their degeneracy. Solid lines correspond to $G_0$ = 1 and dotted lines to $\chi = 10$.}
\label{Fig:ChemForm}
\end{figure}

Figure \ref{Fig:ChemForm} shows the  $N(3,3)$/$N(1,1)$ ratios as a function of $\zeta$ for the three prescriptions. Results are presented for PDR models with \nh~= 100 cm$^{-3}$, $A_\textrm{V}^{max}$~= 1, and $G_0$ = 1 and 10. This last value of $G_0$ corresponds to the order of magnitude of the UV radiation field in the CMZ \citep{Porter05, Moskalenko06}. In these models, \cite{Roncero12} collision rates are used.\\

The $N(3,3)$/$N(1,1)$ ratio exhibits a significant increase for low values of $\zeta$ when scenario A is introduced and when $G_0$ = 1. These conditions correspond to a mean kinetic temperature of $\sim$ 65 K for $\zeta$ = 10$^{-16}$ s$^{-1}$ so that collisions are not efficient to excite the (3,3) level. However, when $G_0$ = 10 under the same low values of $\zeta$, the mean kinetic temperature increases (T reaches $\sim$ 160 K for $\zeta$ = 10$^{-16}$ s$^{-1}$), such that the $N(3,3)$/$N(1,1)$ ratio also increases significantly. Hence, the additional excitation introduced in scenario A is moderate compared to that in scenarios B and C.\\

At high $\zeta$, $N(3,3)$/$N(1,1)$ ratios increase with $\zeta$. In scenario A, the ratio converges toward 14/3, the ratio of the statistical weights of the (3,3) and (1,1) levels since characteristic timescales for statistical equilibrium are controlled by chemical processes. Models B, with the hypothesis that H$_3^+$ is formed following a Boltzmann distribution at gas temperature, also converge toward this value. Indeed, the mean gas temperature is then 360 K and 420 K for, respectively, $G_0$ = 1 and 10, and for $\zeta$ = 10$^{-13}$ s$^{-1}$, so the Boltzmann distribution at gas temperature tends toward a repartition following statistical weights. Since the gas temperature is high and collisional excitation is efficient, even scenario C gives large values of $N(3,3)$/$N(1,1)$. Nevertheless, for this scenario, when $\zeta$ $>$ 10$^{-13}$ s$^{-1}$, $N(3,3)$/$N(1,1)$ decreases with $\zeta$ because collisional excitations no longer have time to contribute to the excitation from low levels (1,1) and (1,0) before the molecule is destroyed.\\

These results show the importance of chemical excitation to compute \hhhp~excitation. Fortuitously, this contribution is not as decisive in CMZ conditions ($G_0$ $\simeq$ 10 and high flux of cosmic rays) as in typical diffuse clouds because gas temperature becomes high enough to enhance collisional excitations. Values close to 1 have been found for the $N(3,3)$/$N(1,1)$ ratio in CMZ observations. These values are obtained in Fig. \ref{Fig:ChemForm} for high values of $\zeta$.  This is discussed in Sect. \ref{Sec:CMZ}. In the following, scenario A is always used except when explicitly said. 

\subsection{Sensitivity to collisional excitation rates}
\label{SubSec:Collision}

Collisional excitation of \hhhp~is a significant source of uncertainty in describing the population of excited levels. We check the sensitivity to the available collision rates by running 
two sets of 176 isothermal and isochoric PDR models, one set with \cite{Oka04} prescription and another one with \cite{Roncero12} collision rates. 
The proton density is varied from 1 to 10$^5$ \cmt~and the temperature range is between 50 and 1000 K. The Galactic values reported in Tables \ref{Tab:InputParam} and \ref{Tab:ElemAbund} are  used 
($G_0$ = 1, $A_\textrm{V}^{max}$ = 1) with a cosmic-ray ionization rate value  $\zeta$ = 10$^{-16}$ s$^{-1}$.  We also assume that 33\% of the exothermicity of the formation reaction excites \hhhp~at formation following scenario A described in Sect. \ref{Sec:ExChim}.\\

\begin{figure*}[t!]
\centering
\includegraphics[width=16.4cm]{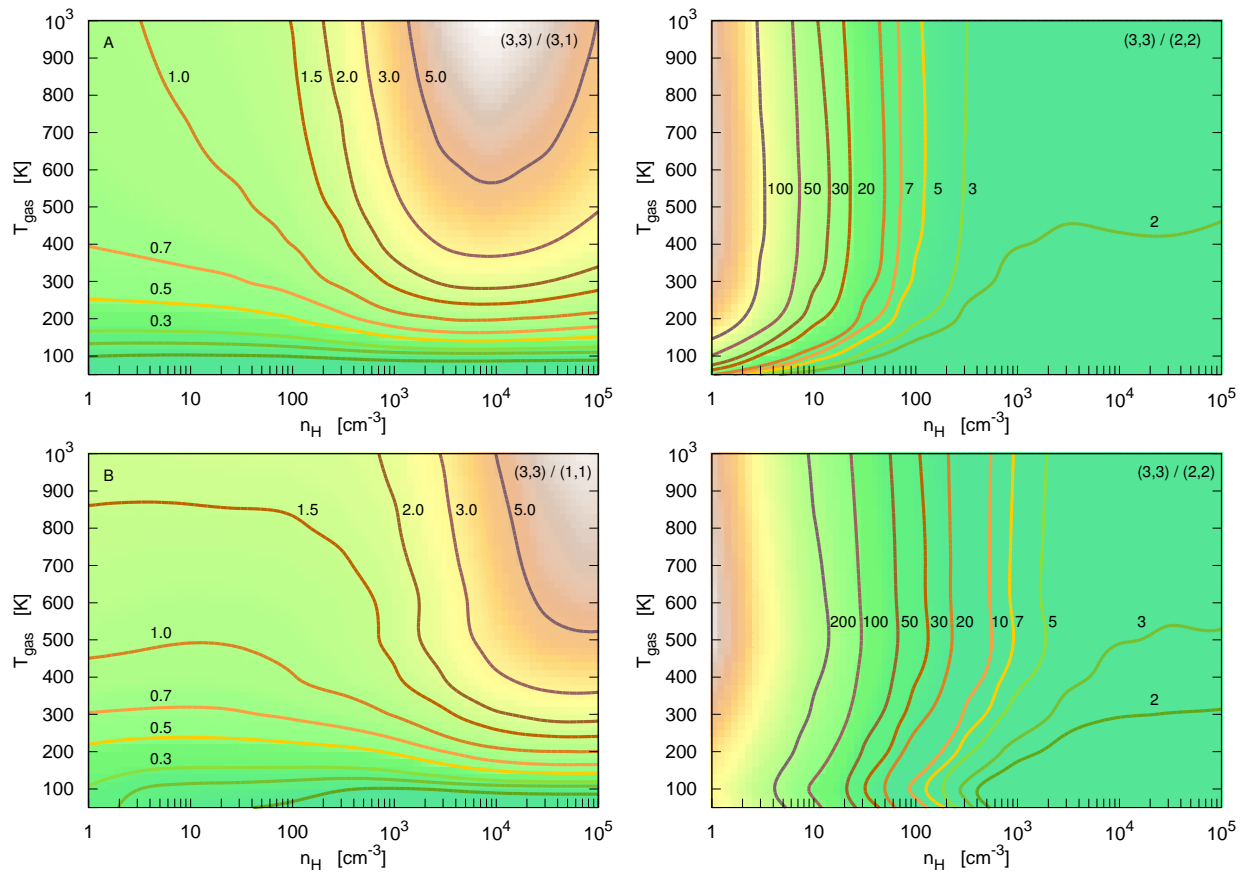}
\caption{Ratios $N(3,3)$/$N(1,1)$ and $N(3,3)$/$N(2,2)$ in the plane \nh-$T$ obtained with the Meudon PDR code. Top (A): models with \cite{Oka04} \hhhp~collision rates. Bottom (B): models with \cite{Roncero12} collision rates.}
\label{Fig:Grid_nH_T}
\end{figure*}

\begin{figure}[h!]
\includegraphics[width=9.2cm]{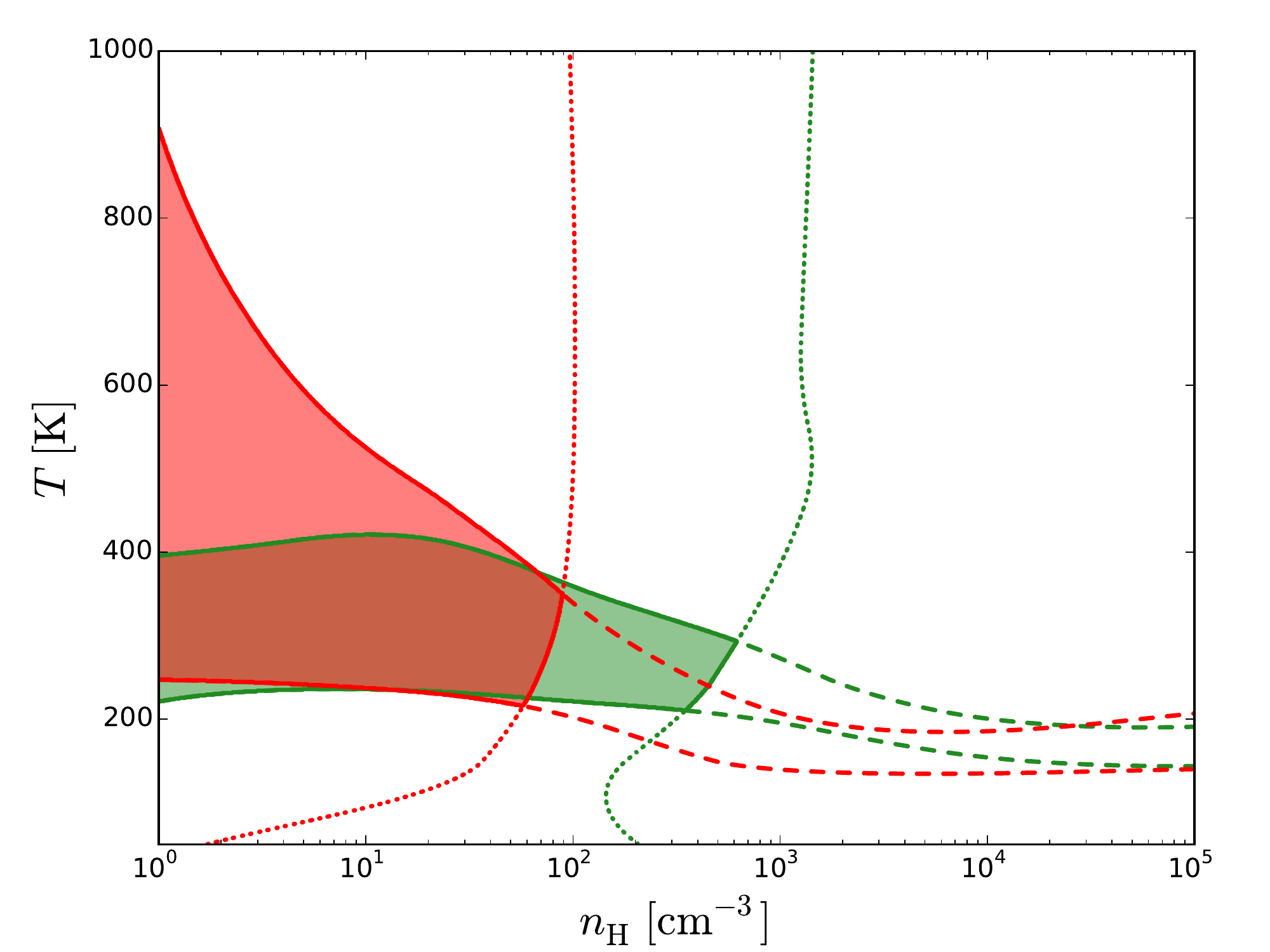}
\caption{$N$(3,3)/$N$(1,1) (dashed lines) and $N$(3,3)/$N$(2,2) (dotted lines) ratios in a plane \nh-$T$ matching observations toward 2MASS J17470898-2829561, $\Delta v $ between -150 and -60 km s$^{-1}$ \citep{Goto11}. Red corresponds to models obtained with \cite{Oka04} collision rates and green to models obtained with \cite{Roncero12} collision rates}
\label{Fig:Grid_nH_T_Obs}
\end{figure}

Figure \ref{Fig:Grid_nH_T} shows corresponding results as contour maps of the ratios $N(3,3)$/$N(1,1)$ and $N(3,3)$/$N(2,2)$. First, we obtain similar results to those reported in \cite{Oka04} and \cite{Oka05} when we use their collision rates prescription, even though our models include chemical excitation. Second, we also find that models including \cite{Roncero12} collision rates suggest slightly higher densities and temperatures than those obtained with \cite{Oka04} prescription. This is more apparent in Fig. \ref{Fig:Grid_nH_T_Obs}, which  presents the space parameter where $N(3,3)$/$N(1,1)$ = 0.57 $\pm$ 0.11 and $N(3,3)$/$N(2,2)$ $>$ 3.21, corresponding to the 2MASS J17470898-2829561 line of sight \citep{Goto11}. We then conclude that, to the first order, the available collisional excitation rates of \hhhp~\citep{Oka04,Roncero12} lead to similar estimates of the density and temperature. In the following, we use \cite{Roncero12} collision rates, which are  are calculated from a precise intermolecular potential surface and take nuclear spin restriction rules into account .\\

\begin{figure*}[t!]
\centering
\includegraphics[width=16.4cm]{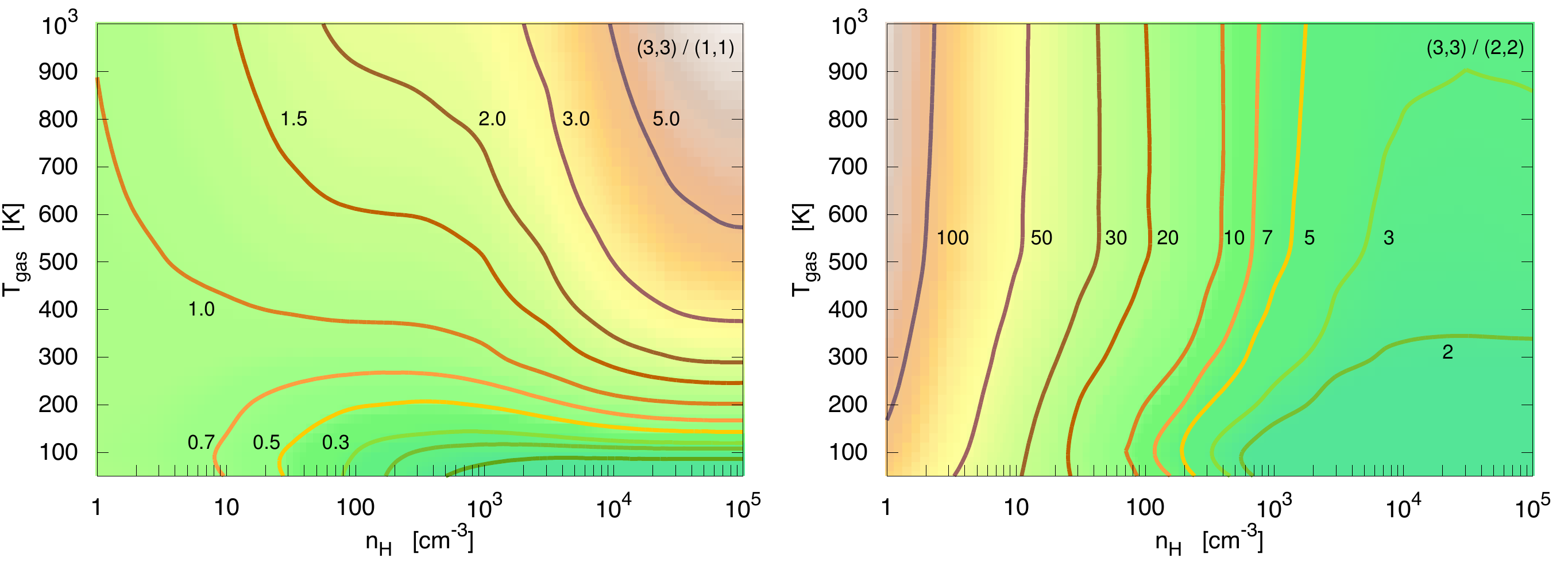}
\caption{Same as bottom panels of Fig. \ref{Fig:Grid_nH_T}, but produced with models with $\zeta = 3\times 10^{-14}$ s$^{-1}$.}
\label{Fig:Grid_nH_T_highZ}
\end{figure*}

Figure \ref{Fig:Grid_nH_T_highZ} presents the same ratios 
 for models with a larger value of  $\zeta$, $3\times 10^{-14}$ s$^{-1}$, which is more appropriate  for CMZ conditions, as emphasized later. Compared to the previous contour maps, the increase of the cosmic-ray flux enhances the $N$(3,3)/$N$(1,1) ratio in the low density ($n_\textrm{H}<100$ \cmt) and low temperature ($T$ < 300 K)  domain because of the faster formation-destruction cycle of H$_3^+$.
 
\subsection{Excitation by IR pumping}
\label{SubSec:IRPumping}

\cite{Goto08} reported a suggestion by John Black: (3,3) level population may be enhanced by IR pumping (see Fig. \ref{Fig:LevelsH3p}). Photons at 3.6685 $\mu$m pump the ortho (1,0) level of the ground vibrational state toward the vibrationally $\nu_2$ = 1 excited state (2,0). This is followed by rapid radiative de-excitation into (3,0) in $2.53\times10^{-2}$ s. Then level (3,0) de-excites toward the (3,3) metastable level through a semiforbidden radiative transition in less than four hours. \cite{Goto08} estimate that, despite an enhanced IR radiation field in the CMZ compared to local conditions, this process should not be competitive with collisional excitation in the CMZ (see their Sect. 5.5). \\

To examine the impact of this process, we increase the number of \hhhp~levels considered in the PDR code up to level $v = 1, J = 2, K=0$.  We use \cite{Roncero12} collision rates for the 24 first levels of \hhhp~and Oka \& Epp prescription for upper levels. Following \cite{Porter05} and \cite{Moskalenko06}, the radiation field in the CMZ at 3.7 $\mu$m could be 20 times higher than in local ISM. In the Meudon PDR code, the intensity of the incident radiation field in the visible and  infrared is modeled by two black bodies: one corresponding to stars emission and one to dust emission. We scale the intensities of these black bodies so that the intensity in the visible - IR impinging on the slab of gas is 20 times the standard radiation field at the wavelength of the R(1,0) transition. \\

We ran several models of diffuse clouds with \nh~ = 100 \cmt, $A_\textrm{V}^{max}$ = 1 and with different values of  $G_0$, $\zeta$ and incident intensity at 3.6685 $\mu$m to cover local and CMZ conditions. Table \ref{Tab:IR} presents $N(3,3)$/$N(1,1)$ and $N(3,3)$/$N(2,2)$ for these models. We find that IR pumping always has negligible impact on the $N$(3,3)/$N$(1,1) ratio. It has a minor effect on the ratio $N$(3,3)/$N$(2,2) in standard conditions. In CMZ conditions, $G_0$ = 10 and $\zeta$ = 10$^{-13}$ s$^{-1}$, IR pumping has no effect compared to the other excitation processes (direct collisions and excitation at formation). These models confirm the statement of \cite{Goto08}. In the following of the paper, we neglect this process and use models with 24 \hhhp~levels, allowing us to use consistent collision rates by \cite{Roncero12}.

\begin{table}[h!]
\center
\caption{Effect of IR pumping on $N(3,3)$/$N(1,1)$ and on $N(3,3)$/$N(2,2)$. The first three columns correspond to input parameters. IR is the scaling factor to the IR part of the ISRF.}
\label{Tab:IR}
\begin{tabular}{lllll}
\hline
\hline
$G_0$    &  $\zeta$        &  IR             & $N(3,3)$/$N(1,1)$  &  $N(3,3)$/$N(2,2)$ \\
\hline
1        &  $10^{-16}$     & $\times$ 1      & 0.1          &  7.5   \\
1        &  $10^{-16}$     & $\times$ 20     & 0.1          &  5.9   \\
1        &  $10^{-13}$     & $\times$ 1      & 1.0          &  9.4   \\
1        &  $10^{-13}$     & $\times$ 20     & 0.9          &  7.7   \\
10       &  $10^{-16}$     & $\times$ 1      & 0.4          &  8.6   \\
10       &  $10^{-16}$     & $\times$ 20     & 0.4          &  8.1   \\
10       &  $10^{-13}$     & $\times$ 1      & 1.2          &  8.6   \\
10       &  $10^{-13}$     & $\times$ 20     & 1.2          &  8.2   \\
\hline
\end{tabular}
\end{table}

\section{Physical conditions in the CMZ}
\label{Sec:CMZ}

In this section, we try to constrain the physical conditions and processes in the CMZ, including the cosmic-ray ionization rate. We ran a grid of 1776 isochoric PDR models with parameters relevant to the CMZ conditions. The metallicity, Z, in the CMZ is between 2 and 5 times higher than in the solar neighborhood \citep{Rolleston00, Daflon04, Rudolph06}. We set it to Z = 3. This increases elemental and  grain abundances as described in Tables \ref{Tab:InputParam} and \ref{Tab:ElemAbund}. In all models, the incident radiation field in the far-UV impinging on both sides of the cloud is the ISRF scaled by a factor\footnote{In Sect. \ref{SubSec:IRPumping}, we tested with the maximal possible value, $G_0$ = 20 to study the possible impact of IR pumping. Here, we use an intermediate value, $G_0$ = 10. This has no impact on the results.} $G_0$ = 10 \citep{Porter05}. The grid of models is divided in five sets of proton densities, \nh~=~10, 20, 50, 100, 1000 \cmt. For each proton density, $\zeta$ ranges from $10^{-17}$ to $10^{-12}$ s$^{-1}$, and the size of the clouds, $A_\textrm{V}^{max}$, ranges from 0.2 to 6. We take the full physics described above  into account except for the enhanced IR pumping of H$_3^+$, since we found that this process has a negligible effect on \hhhp~excitation. This simplification allows us to limit the computation of \hhhp~excitation to its 24 first levels for which collision rates are reliable. In all models, the gas temperature is computed at each position of the cloud taking  heating and cooling mechanisms implemented in the Meudon PDR code into account. For models with constant density, the relationship between the size in cm, $L$, and the visual extinction, $A_\textrm{V}^{max}$, is $L = 6.1 \,\, A_{\textrm{V}}^{max} / Z \, \left(100 / n_{\textrm{H}} \right)$ pc.

\subsection{Estimates of $\zeta$ and L with H$_3^+$}

To infer physical conditions on each line of sight where \hhhp~is detected, we use a $\chi^2$ minimization built with \hhhp~column densities in levels (1,1), (3,3), and (1,0) when known. We also check that the best models according to the minimization procedure are compatible with the upper limits on the (2,2) level. Table \ref{Tab:Obs} summarizes \hhhp~observed column densities on the 10 CMZ lines of sight on which $\chi^2$ minimizations are performed.\\

\begin{table}[h!]
\center
\caption{Observed column densities of H$_3^+$ in the CMZ in  (1,1), (1,0), (3,3), and (2,2) levels. Column densities are expressed in 10$^{14}$ cm$^{-2}$.}
\label{Tab:Obs}
\begin{tabular}{lllll}
\hline
\hline

Source      &  $N(1,1)$       &  $N(1,0)$    &  $N(3,3)$       & $N(2,2)$\\
\hline
GC IRS 21   &  28.1$\pm$12.3  &  9.0$\pm$4.4 &  24.2$\pm$12.1  & < 8.3   \\
GC IRS 3    &  10.8$\pm$2.1   &  3.4$\pm$1.3 &   8.4$\pm$1.9   & < 1.6   \\
GC IRS 1W   &  18.1$\pm$3.8   &  7.9$\pm$2.4 &  11.7$\pm$3.0   & < 3.3   \\
GCS 3-2     &  17.0$\pm$1.7   &  4.6$\pm$0.8 &   9.8$\pm$1.6   & < 3.0   \\
J1743       &  12.8$\pm$1.9   &  -           &   5.7$\pm$2.2   & < 2.1   \\
J1747       &   6.0$\pm$0.7   &  -           &   4.3$\pm$1.3   & < 0.8   \\
NHS 21      &   8.9$\pm$2.2   &  3.7$\pm$1.5 &   5.6$\pm$1.4   & < 2.3   \\
NHS 22      &  16.9$\pm$5.6   &  7.9$\pm$3.5 &   9.7$\pm$2.7   & < 3.6   \\
NHS 25      &  11.4$\pm$5.9   &  4.0$\pm$3.8 &   7.4$\pm$5.3   & < 3.9   \\
NHS 42      &  17.7$\pm$5.1   &  8.4$\pm$4.1 &   8.6$\pm$4.1   & < 4.0   \\
Mean        &  14.5$\pm$0.8   &  6.5$\pm$1.1 &   8.6$\pm$1.0   &         \\
\hline
\end{tabular}
\tablebib{J1747 corresponds to 2MASS J17470898-2829561 and J1743 to 2MASS J17432173-2951430. Data are from \cite{Goto08} for all  lines of sight except for J1747 and J1743 for which data come from \cite{Goto11}. The GCS 3-2 line of sight is not taken into account to compute the weighted mean value (see text).}
\end{table}

\begin{figure*}[t!]
\includegraphics[width=18.4cm]{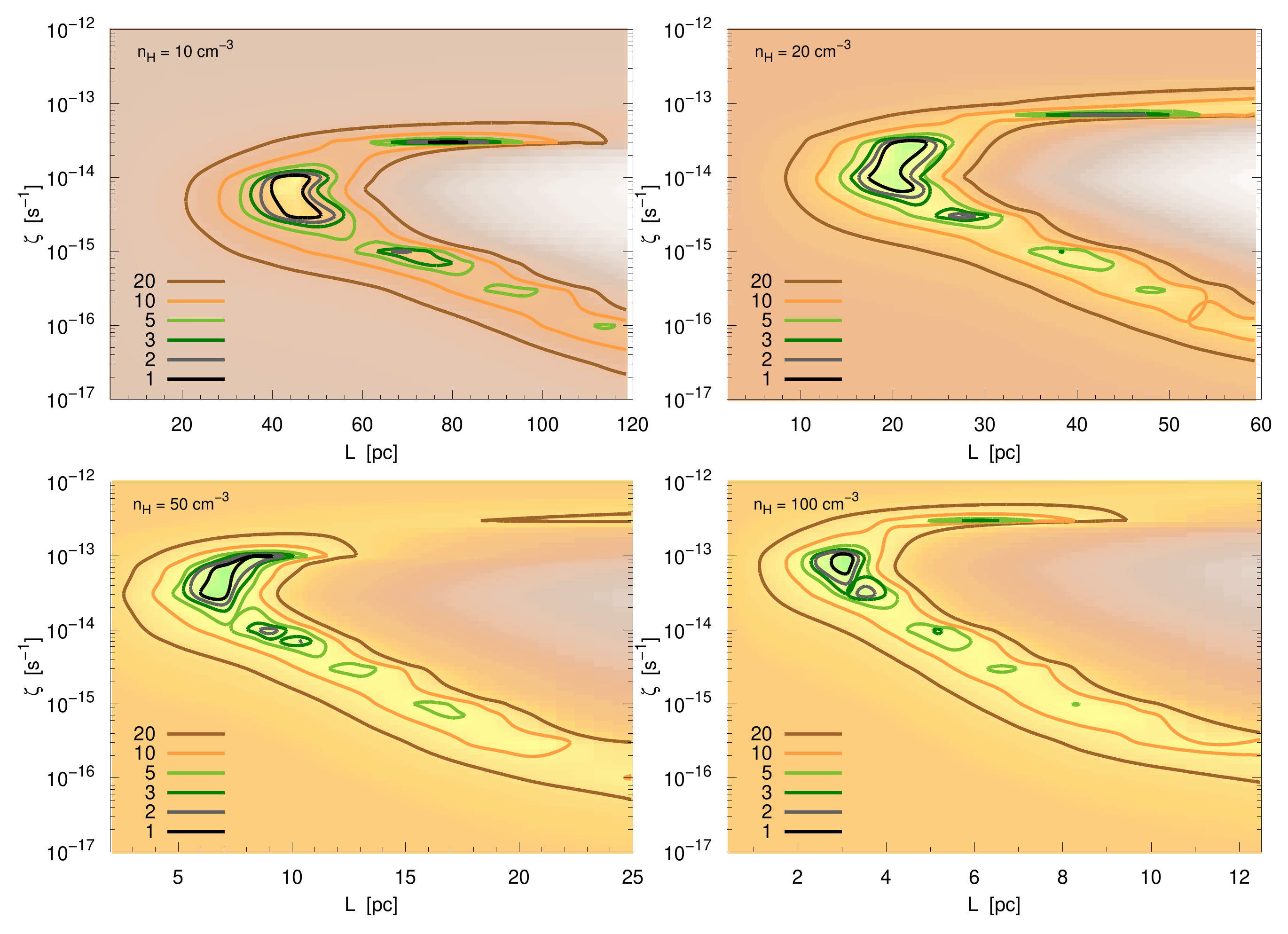}
\caption{Contour map of  $\chi^2$  using H$_3^+$ column densities $N$(1,1) and $N$(3,3) toward J1747 for the four densities.} 
\label{Fig:Chi2H3p}
\end{figure*}

\begin{figure}[h!]
\includegraphics[width=9.2cm]{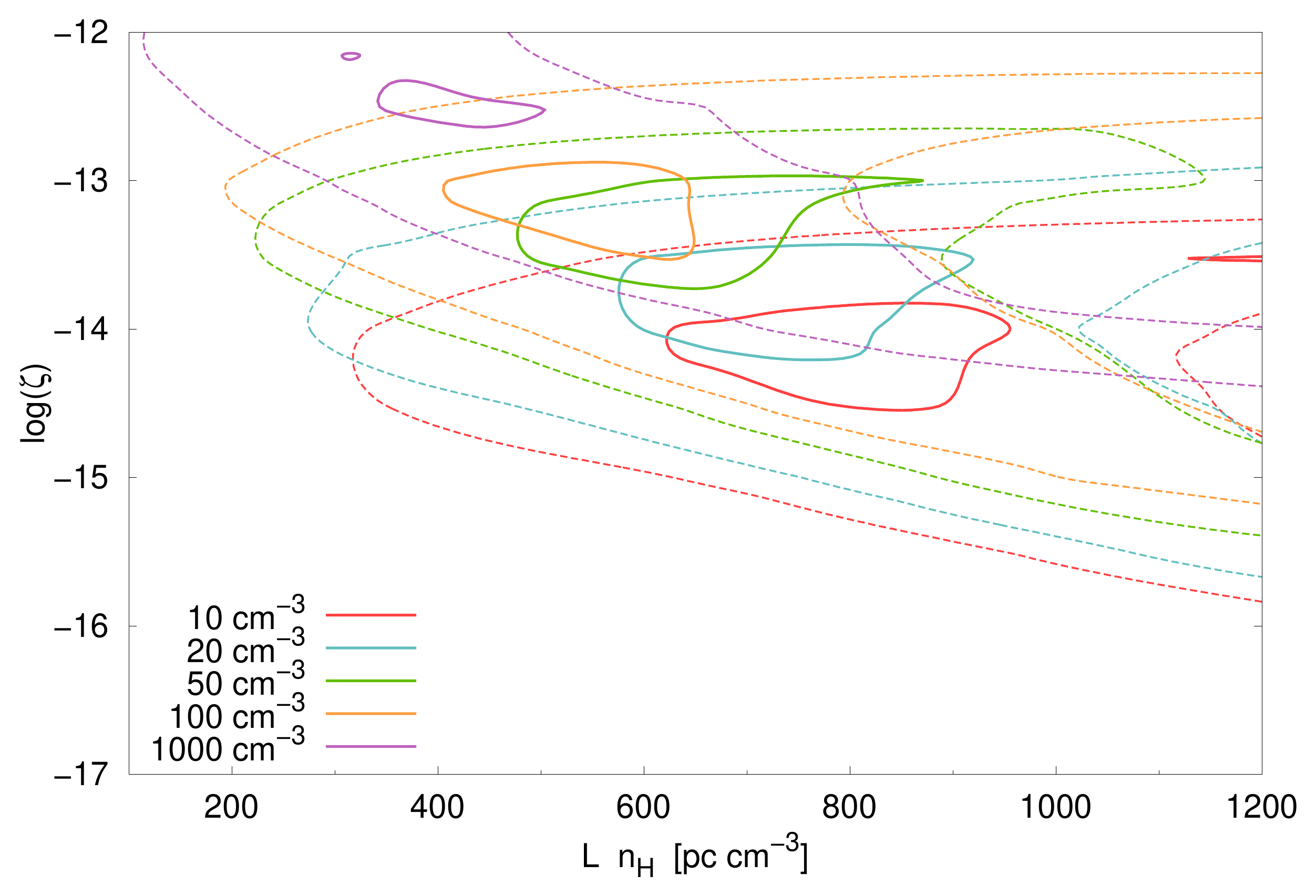}
\caption{Comparison of the $\chi^2$ contours for the five densities for the NHS 22 line of sight. Solid lines correspond to $\chi^2$ = 1 and dashed lines to $\chi^2$ = 5.}
\label{Fig:Chi2ComparenH}
\end{figure}

For all lines of sight, except GCS 3-2, a clear $\chi^2 < 1$ zone is found with models \nh~=~10 to 100 \cmt~(Fig. \ref{Fig:Chi2H3p} shows an illustration of the $\chi^2$ minimization for the J1747 line of sight). Models with \nh~=1000 \cmt~ only reach a $\chi^2 < 1$  for GCS IRS 21, NHS 22, and, marginally, in the case of J1747. Moreover, when a $\chi^2$ of 1 is reached for \nh~=1000 \cmt~ models, column densities in the (2,2) level are only lower than the upper limit  by a few percent. For these reasons, even if it is formally not possible to reject models with densities of 1000 \cmt, this density is less likely. As expected, we find that the lower the density, the lower the cosmic-ray ionization rate and the larger the size of the cloud. Figure \ref{Fig:Chi2ComparenH} presents a comparison of $\chi^2$ contours for the five densities for the line-of-sight NHS 22.\\

Table \ref{Tab:chi2} lists, for each line of sight, the positions in the $\zeta-L$ plane where $\chi^2$ values are minimal, as well as the corresponding computed \hhhp~level column densities, molecular fractions,  average temperatures, and electronic fractions. Assuming \hhhp~has similar properties on the various lines of sight as demonstrated by \cite{Goto05, Goto08, Goto11}, we can derive a weighted mean of observed column densities (Table \ref{Tab:Obs}). This weighted mean is computed assuming the uncertainties in observations are dominated by photon noise and therefore, follow a Poisson process. We exclude from this weighted mean the GCS 3-2 line of sight since it seems peculiar. A $\chi^2$ minimization on this weighted mean leads to a cosmic-ray ionization rate and a path length ranging from $10^{-14}$ s$^{-1}$ and 66 pc for \nh~= 10 \cmt, to $11 \times 10^{-14}$ s$^{-1}$ and 4 pc for \nh~=~100 \cmt (see Table \ref{Tab:chi2}). It is difficult to constrain  $\zeta$ and $L$ further without additional information about the gas density.\\

Our value of $\zeta$ is somewhat larger than the estimate by \cite{Yusef13}, who deduced from synchrotron and Fe K$\alpha$ observations, $\zeta_1 \sim 10^{-15} - 10^{-14}$ s$^{-1}$. However, this value is not incompatible since low-energy cosmic-ray ions may contribute to ionization without producing synchrotron emission and, according to \cite{Strong10}, the luminosity of cosmic-ray protons is $\sim$70 times that of cosmic-ray electrons.\\



\begin{table*}[h!]
\center
\small
\caption{Results of the best models defined as those minimizing $\chi^2$ for the various densities, $n_\textrm{H}$ = 10, 20, 50, and 100 cm$^{-3}$. Cosmic-ray ionization rates, $\zeta$, are in 10$^{-14}$ s$^{-1}$ and sizes, $L$, in parsecs. Computed \hhhp~column densities, $N(J,K)$, are in 10$^{14}$ cm$^{-2}$. The  last three columns present computed molecular fractions, mean gas temperatures in Kelvin, and mean electronic fractions in 10$^{-3}$.}
\label{Tab:chi2}
\begin{tabular}{l l l l l l l l l l l}
\hline
\hline
\multicolumn{11}{c}{$n_\textrm{H}$ = 10 cm$^{-3}$} \\
\hline
Source      &  $\zeta$  &  $L$    &  $\chi^2$ &  N(1,1)  &  N(1,0) &  N(3,3) & N(2,2)    & f     & $<\!\!T\!\!>$    &  $<\!\!x_e\!\!>$\\
\hline
GC IRS 21   &   1       &   94    &  0.6      &  19.5   &  11.5  &  12.8  &   0.14   & 0.5   &  316 &  1.3     \\
GC IRS 3    &   3       &   108   &  0.3      &  9.6    &  4.4   &  9.4   &   0.11   & 0.4   &  471 &  3.0     \\
GC IRS 1W   &   1       &   81    &  0.3      &  15.8   &  9.2   &  10.4  &   0.11   & 0.5   &  321 &  1.4     \\
GCS 3-2     &   3       &   121   &  4.79     &  11.0   &  5.0   &  10.9  &   0.13   & 0.4   &  467 &  3.0     \\
J1743       &   0.3     &   74    &  0.006    &  13.0   &  8.8   &  5.0   &   0.05   & 0.7   &  212 &  0.6     \\
J1747       &   1       &   45    &  0.004    &  5.8    &  3.1   &  4.3   &   0.05   & 0.5   &  391 &  1.7     \\
NHS 21      &   1       &   53    &  0.2      &  7.9    &  4.4   &  5.5   &   0.06   & 0.5   &  354 &  1.5     \\
NHS 22      &   1       &   78    &  0.08     &  14.8   &  8.6   &  9.7   &   0.11   & 0.5   &  323 &  1.4     \\
NHS 25      &   3       &   108   &  0.08     &  9.6    &  4.4   &  9.5   &   0.11   & 0.3   &  474 &  3.0     \\
NHS 42      &   0.7     &   76    &  0.1      &  15.8   &  9.9   &  8.6   &   0.09   & 0.6   &  278 &  1.0     \\
Mean        &   1       &   66    &  0.5      &  11.2   &  6.4   &  7.5   &   0.08   & 0.5   &  337 &  1.4     \\
\hline
\multicolumn{11}{c}{$n_\textrm{H}$ = 20 cm$^{-3}$} \\
\hline
Source      &  $\zeta$  &  $L$    &  $\chi^2$ &  N(1,1)  &  N(1,0) &  N(3,3) & N(2,2)    & f     & $<\!\!T\!\!>$    &  $<\!\!x_e\!\!>$\\
\hline
GC IRS 21   &  3     &   51   &  0.37    &  22.0    &  12.6     &  17.0     &  0.35  & 0.5   & 319   &  1.3     \\
GC IRS 3    &  6     &   61   &  0.46    &  10.1    &  4.5      &  11.1     &  0.25  & 0.4   & 472   &  3.0     \\
GC IRS 1W   &  3     &   41   &  0.19    &  17.0    &  9.8      &  12.9     &  0.26  & 0.5   & 315   &  1.3     \\
GCS 3-2     &  3     &   34   &  4.90    &  12.8    &  7.3      &  9.7      &  0.20  & 0.5   & 320   &  1.4     \\
J1743       &  0.6   &   33   &  0.01    &  12.5    &  8.6      &  4.9      &  0.10  & 0.7   & 187   &  0.6     \\
J1747       &  3     &   22   &  0.21    &  6.6     &  3.6      &  5.2      &  0.11  & 0.5   & 337   &  1.6     \\
NHS 21      &  2     &   25   &  0.17    &  8.4     &  4.7      &  6.3      &  0.13  & 0.5   & 320   &  1.4     \\
NHS 22      &  2     &   37   &  0.15    &  15.8    &  9.3      &  11.1     &  0.23  & 0.6   & 299   &  1.2     \\
NHS 25      &  3     &   29   &  0.07    &  10.4    &  5.9      &  7.9      &  0.17  & 0.5   & 321   &  1.4     \\
NHS 42      &  1     &   35   &  0.14    &  16.2    &  10.5     &  8.7      &  0.18  & 0.6   & 238   &  0.8     \\
Mean        &  3     &   33   &  0.08    &  12.2    &  6.9      &  9.3      &  0.19  & 0.5   & 321   &  1.4     \\
\hline
\multicolumn{11}{c}{$n_\textrm{H}$ = 50 cm$^{-3}$} \\
\hline
Source      &  $\zeta$  &  $L$  &  $\chi^2$ &  N(1,1)  &  N(1,0) &  N(3,3) & N(2,2)   & f   & $<\!\!T\!\!>$ & $<\!\!x_e\!\!>$\\
\hline
GC IRS 21   &12         &24    &0.1 &16.0  &8.0   &17.7   &0.84   & 0.4   &  432  &  2.1     \\
GC IRS 3    &12         &12    &0.5 & 7.0  &3.5   & 7.7   &0.37   & 0.4   &  424  &  2.2     \\
GC IRS 1W   &6          &14    &0.2 &16.5  &9.9   &13.1   &0.59   & 0.6   &  302  &  1.1     \\
GCS 3-2     &12         &15    &4.4 & 9.4  &4.7   &10.4   &0.50   & 0.4   &  426  &  2.1     \\
J1743       &3          &10    &0.2 & 12.1 &8.0   & 6.6   &0.31   & 0.7   &  220  &  0.7     \\
J1747       &5          &7     &0.1 & 6.4  &3.8   & 4.9   &0.23   & 0.6   &  295  &  1.2     \\
NHS 21      &5          &8     &0.2 & 7.9  &4.7   & 6.1   &0.28   & 0.6   &  295  &  1.2     \\
NHS 22      &5          &12    &0.2 &14.9  &9.2   &10.7   &0.49   & 0.6   &  278  &  1.0     \\
NHS 25      &6          &9     &0.1 & 9.9  &5.8   & 7.8   &0.36   & 0.5   &  303  &  1.2     \\
NHS 42      &3          &12    &0.1 &15.6  &1.0   & 8.8   &0.41   & 0.7   &  227  &  0.7     \\
Mean        &6          &11    &0.1 &12.1  &7.2   & 9.5   &0.44   & 0.6   &  301  &  1.1     \\
\hline
\multicolumn{11}{c}{$n_\textrm{H}$ = 100 cm$^{-3}$} \\
\hline
Source      &  $\zeta$  &  $L$   &  $\chi^2$ &  N(1,1)  &  N(1,0) &  N(3,3) & N(2,2)  & f   & $<\!\!T\!\!>$ & $<\!\!x_e\!\!>$\\
\hline
GC IRS 21   &11   &7   &0.5  &18.6   &11.4   &15.9   & 1.3   &  0.6  &  310  &  1.0     \\
GC IRS 3    &30   &8   &1.1  & 7.7   & 3.7   &10.2   &0.88   &  0.4  &  505  &  2.6     \\
GC IRS 1W   &10   &6   &0.4  &14.8   & 9.3   &12.0   &0.97   &  0.6  &  294  &  1.0     \\
GCS 3-2     &11   &4   &7.0  &11.0   & 6.8   & 9.0   &0.74   &  0.6  &  297  &  1.0     \\
J1743       &3    &5   &0.04 &12.9   & 9.7   & 4.9   &0.44   &  0.7  &  160  &  0.4     \\
J1747       &9    &3   &0.2  & 6.1   & 3.8   & 4.7   &0.40   &  0.6  &  277  &  1.0     \\
NHS 21      &10   &3   &0.3  & 7.4   & 4.6   & 5.9   &0.50   &  0.6  &  286  &  1.0     \\
NHS 22      &8    &5   &0.1  &14.1   & 9.2   & 9.8   &0.80   &  0.6  &  256  &  0.8     \\
NHS 25      &11   &4   &0.1  & 8.9   & 5.5   & 7.3   &0.60   &  0.6  &  293  &  1.0     \\
NHS 42      &6    &5   &0.2  &15.1   &10.2   & 9.3   &0.77   &  0.7  &  233  &  0.7     \\
Mean        &11   &4   &0.07 &10.5   & 6.4   & 8.9   &0.73   &  0.6  &  307  &  1.1     \\
\hline
\end{tabular}
\end{table*}

\subsection{Additional constraints: OH${^+}$, H$_2$O$^+$, H$_3$O$^+$, and HF}
\label{Sec:OHp}
\begin{figure}[h!]
\includegraphics[width=9.2cm]{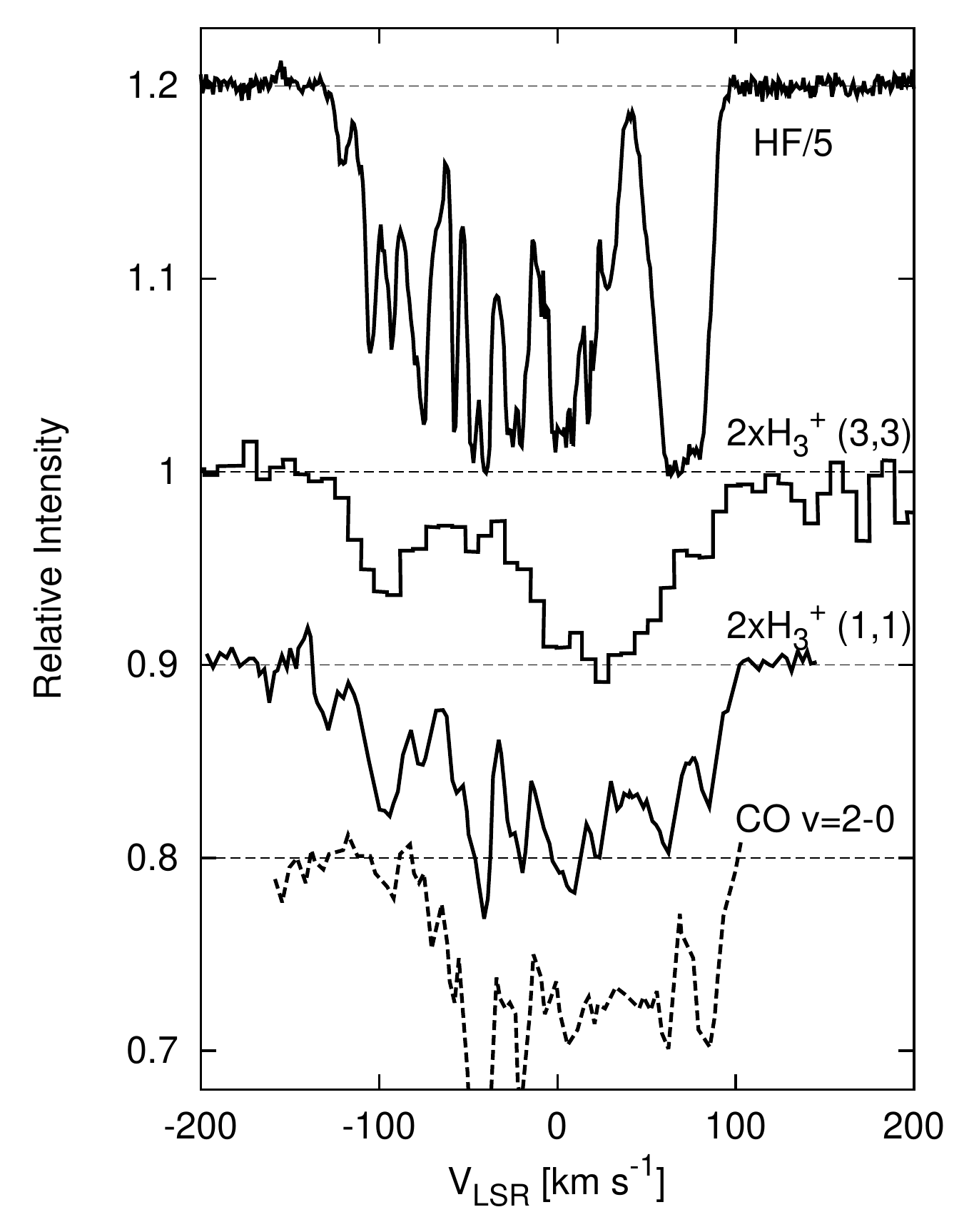}
\caption{HF spectrum for the J$=$1-0 transition obtained with Herschel/HIFI toward Sgr B2(N) (HEXOS key program) and  \hhhp~spectrum for the R(3,3) transition \citep{Goto11} and for the R(1,1)$^l$ \citep{Geballe10} toward J1747. The \hhhp~spectrum of the R(3,3) transition has been corrected by the CO absorption. See Sect. 2 in \cite{Goto11} for more details.}
\label{Fig:Obs_HF}
\end{figure}

Thanks to PRISMAS and HEXOS Herschel key programs, several hydride molecules (HF, CH, OH$^+$, H$_2$O$^+$, H$_3$O$^+$, CH$^+$, SH$^+$, ...) have been observed in absorption toward several lines of sight in the CMZ \citep{Gerin10a, Gerin10b, Neufeld10, Godard12}, but not on the lines of sight where \hhhp~is detected. Nevertheless, the Sgr B2(N) line of sight, part of the HEXOS program, is close to J1747. The projected distance between the two objects is 18 pc (assuming a distance to the Galactic center of 8 kpc). \cite{Oka15} noticed a strong similarity between H$_2$O$^+$ velocity components toward Sgr B2(M) and those of H$_3^+$ toward J1747, and \cite{Indriolo15} observe (see their Fig. 4 and 5) similar components between Sgr B2(N) and Sgr B2(M). Figure \ref{Fig:Obs_HF} presents HF and \hhhp~spectra toward Sgr B2 (N) and J1747. Both molecules present similar velocity absorption components. Absorption features associated with the diffuse gas in the CMZ are located between -150 and -60 km s$^{-1}$ \citep{Goto11}. In this velocity range, HF presents four absorption components, H$_3^+$ one unresolved (3,3) absorption feature and 3 (1,1) components. In addition, weak CO absorption components are detected. Concerning the other absorption components, the strong absorption of HF at \vlsr~ = 60 \kms~is related to the Sgr B2(N) complex. Three absorption components due to spiral arms are present at -40, -20, and 0 \kms, deduced from a longitude-velocity map of the CO J$=$1-0 emission in this region \citep{Oka12}.\\

These similarities in the velocity components between -150 and -60 km s$^{-1}$ may indicate that the size of the diffuse component containing \hhhp~, toward J1747, extends up to the Sgr B2(N) line of sight and, therefore, the size of this component should be at least of $\sim$ 18 pc. This is also compatible with the fact that \hhhp~is largely observed in the CMZ with always the same peculiar properties, indicating that the filling factor of the neutral diffuse gas probed by \hhhp~must be large. According to Table \ref{Tab:chi2}, this leads to \nh~$< 50$ cm$^{-3}$ for the J1747 line of sight. As a consequence, the cosmic-ray ionization rate should be $\zeta$ < 5$\times10^{-14}$ s$^{-1}$.\\

Assuming that observations toward J1747 and Sgr B2(N) probe the same diffuse gas, derived column densities toward Sgr B2(N) can be used to test our models. We reanalyzed Herschel HF observations. Column densities in the four components between -150 and -60 km s$^{-1}$ are given in Table \ref{Tab:HFcoldens}. As discussed by \cite{Godard12}, $N(\textrm{HF})$ may provide an upper limit to $N(\textrm{H}_2)$. With an elemental abundance F/H of $1.8\times10^{-8}$, the value in the solar neighborhood, and a molecular fraction of 1, \cite{Godard12} find $N$(HF)/$N$(H$_2$) = $3.6\times10^{-8}$ (also reported in \cite{Neufeld05}). This ratio is only valid for media where Z~=~1 and where $n(\textrm{e}^-) = n(\textrm{C}^+)$. To estimate a better ratio in the CMZ, we use Eq. C.3 of \cite{Godard12} without the hypothesis that the density of electrons is given by C$^+$, with Z = 3, and with the gas temperature, electron density, and molecular fraction derived from our best models for J1747 (Table \ref{Tab:chi2}). We find $N(\textrm{HF})/N(\textrm{H}_2) = 1.1\times10^{-7}$. Derived N(H$_2$) are presented in Table \ref{Tab:HFcoldens}. 

\begin{table}[h!]
\center
\caption{Observed HF and estimated \hh~column densities toward Sgr B2(N).}
\label{Tab:HFcoldens}
\centering
\begin{tabular}{rrll}
\hline
\hline
v$_{min}$   & v$_{max}$    & $N(\textrm{HF})$            & $N(\textrm{H}_2)$         \\
(\kms)      & (\kms)       & ($\times 10^{13}$\cmd)      & ($\times 10^{20}$\cmd)         \\
\hline
-150        & -114         & 0.49 $\pm$ 0.05             & 0.4  $\pm$ 0.04         \\
-114        & -100         & 2.14 $\pm$ 0.65             & 1.9  $\pm$ 0.6         \\
-100        & -88          & 2.15 $\pm$ 0.61             & 1.9  $\pm$ 0.5                 \\
-88         & -60          & 5.86 $\pm$ 0.64             & 5.3 $\pm$ 0.6                 \\
\hline
-150        & -60          & 10.4 $\pm$ 1.7              & 9.4  $\pm$ 1.5         \\
\hline
\end{tabular}
\end{table}

The chemistry of HF is included in the Meudon PDR code. Using our grid of models, we tried a $\chi^2$ minimization built with the column densities of \hhhp~in (1,1), (1,0), and (3,3) toward J1747 and $N$(HF) toward Sgr B2(N). None of our models reproduce these four constraints at the same time. When $N$(\hhhp, J, K) are reproduced, $N$(HF) is underpredicted by a factor 4. A possible explanation could be that a fraction of HF observed toward Sgr B2(N) be associated with denser gas than the one probed by \hhhp (but not too dense either because of the lack of CO absorption in the -150 to -60 km s$^{-1}$ velocity range). On the other hand, the elemental abundance of F is not well understood and is a widely discussed topic. We scaled it by Z but it is possible that F/H does not scale as the other elements (see \cite{Abia15} and references therein).\\ 

The observations of OH$^+$, H$_2$O$^+$, and H$_3$O$^+$  are particularly interesting since, like \hhhp, the abundances of these molecules depend on the cosmic-ray ionization rate. \cite{Schilke10,Schilke13} report the detection of the H$_2$O$^+$ ion toward Sgr B2 with Herschel, whereas \cite{Lis12,Lis14} report H$_3$O$^+$ detection. \cite{Indriolo15} derive their column densities in the different velocity components. With a simple analytic relationship between  $N(\textrm{OH}^+)$ and $N(\textrm{H}_2\textrm{O}^+)$, they deduce a value of $\zeta$ $\sim 10^{-14}$ s$^{-1}$, which is compatible with our value if the proton density is $\sim$10 \cmt~and lower than our value if the proton density is higher. The oxygen chemistry is included in the Meudon PDR code, and therefore  we also derive $N(\textrm{OH}^+)$, $N(\textrm{H}_2\textrm{O}^+)$ and $N(\textrm{H}_3\textrm{O}^+)$ together with N(\hhhp). In Table
\ref{Tab:OHp}, we report  these computed column densities for the best models matching H$_3^+$ observations toward J1747, for \nh~= 10, 20, and 50 cm$^{-3}$, and compare them to the observed values toward Sgr B2(N) in the -130, -60 km s$^{-1}$ velocity range. We find a very remarkable agreement. The \nh~= 50 \cmt~model underestimates  the observations only slightly. The computed column density of H$_3$O$^+$ is also compatible with the observations that are reported for the lowest (1,0) ortho level. So, our values of $\zeta$ deduced with H$_3^+$ observations are consistent  with OH$^+$, H$_2$O$^+$ and H$_3$O$^+$ observations as well.\\

A significant discrepancy between our models and \cite{Indriolo15} results concerns the molecular fraction. \cite{Indriolo12a,Indriolo15} estimate the molecular fraction using H$_2$O$^+$ main formation and destruction reactions
\begin{eqnarray*}
\textrm{OH}^+              + \textrm{H}_2    \longrightarrow \textrm{H}_2\textrm{O}^+ + \textrm{H}    &  k_3 = 1.1\times 10^{-9}  \quad \textrm{cm}^3  \textrm{s}^{-1}\\
\textrm{H}_2\textrm{O}^+   + \textrm{H}_2    \longrightarrow \textrm{H}_3\textrm{O}^+ + \textrm{H}    &  k_4 = 6.1\times 10^{-10} \quad \textrm{cm}^3  \textrm{s}^{-1}\\
\textrm{H}_2\textrm{O}^+   + \textrm{e}^-    \longrightarrow products                                 &  k_5 = 4.3\times 10^{-7} (T/300)^{-0.5} \,\, \textrm{cm}^3  \textrm{s}^{-1},\\
\end{eqnarray*}
and they obtain:
\begin{equation}
\label{Eq:Indriolof}
f = \frac{2 \, x_e \, k_5 / k_3}{N(\textrm{OH}^+) / N(\textrm{H}_2\textrm{O}^+) - k_4 / k_3}
.\end{equation}
The use of this expression relies on guesses of the electronic fraction and   gas temperature. Equation \ref{Eq:Indriolof} is used in \cite{Indriolo15} to derive $f$ in 105 diffuse gaseous components and leads to extremely small $f$ values, lower than 0.1 for most components. For Sgr B2(N), in the -130, -60 km s$^{-1}$ velocity component, \cite{Indriolo15} obtain $f$ = 0.08 $\pm$ 0.02 whereas we find $f \sim 0.5-0.6$ for J1747.\\

The molecular fraction may also be directly derived from observations of H at 21 cm and estimate of N(H$_2$) (Table \ref{Tab:HFcoldens}). The column density of H is reported as (1.03 $\pm$ 0.3)$\times$10$^{21}$ cm$^{-2}$ in the -130, -60 km s$^{-1}$ CMZ velocity component of the Sgr B2(N) line of sight by \cite{Indriolo15}, whereas \cite{Godard12} report $N$(H$_2$) = (2.0 $\pm$ 0.5) $\times$ 10$^{21}$ cm$^{-2}$. We previously re-estimate this value to be (9.4 $\pm$ 1.5)$\times$10$^{20}$ cm$^{-2}$. The resulting molecular fraction is then 0.6 (0.8 with the \cite{Godard12} estimate). 
These molecular fractions are then in very good agreement with those computed by the Meudon PDR code.\\

The origin of the discrepancy arises from the estimate of $T$ and $x_e$. \cite{Indriolo15} assume a gas temperature of 70 K and an electronic fraction of 1.5$\times 10^{-4}$, whereas the computed values in the PDR models that match J1747 and Sgr B2(N) observations are $<\!\!x_e\!\!> = $ $1-2\times10^{-3}$ and $<\!\!T\!\!> = $ 295 to 391 K for \nh~from 50 to 10 \cmt. Using these values in Eq. \ref{Eq:Indriolof} leads to f $\simeq$ 0.3-0.6. These conflicting conclusions illustrate our discussion in Sect. \ref{Sec:ChemH3p} in which we highlight the risks of using speculated values in simple chemical analytic expressions,  like  Eqs. \ref{Eq:Indriolo} or \ref{Eq:Indriolof}, especially in exotic conditions such as the Galactic center. 

\begin{table}[h!]
\center
\caption{Comparison of $N(\textrm{OH}^+)$, $N(\textrm{H}_2\textrm{O}^+)$ and $N(\textrm{H}_3\textrm{O}^+)$ observed toward Sgr B2(N) in the -130, -60 km s$^{-1}$ velocity component to the computed values by the Meudon PDR code for the J1747 best models. Column densities for OH$^+$ and H$_2$O$^+$ are in $10^{14}$ cm$^{-2}$ and in $10^{13}$ cm$^{-2}$ for H$_3$O$^+$. Cosmic-ray ionization rate is in $10^{-14}$ s$^{-1}$ and size is in pc.}
\label{Tab:OHp}
\begin{tabular}{l l l l l}
\hline
\hline
                              & Observations$^{(a)}$  &  \multicolumn{3}{c}{Best models for J1747}\\
                              & Sgr B2(N)             &  10 \cmt  &  20 \cmt & 50 \cmt     \\
\hline
$\zeta$                       &                       &  1.0   &  2.6   &  5.0  \\
$L$                           &                       &  45    &  22    &  7    \\
\hline
$N$(OH$^+$)                   &  7.5$\pm$1.5          &  7.1   &  8.2   &  4.1  \\
$N$(H$_2$O$^+$)               &  2.2$\pm$0.4          &  2.3   &  2.4   &  1.6  \\
$N$(H$_3$O$^+$)               &  4.0$\pm$0.8          &  5.1   &  5.4   &  4.9   \\
\hline
\end{tabular}
\tablebib{ (a) \cite{Indriolo15}. The reported value of $N(\textrm{H}_3\textrm{O}^+)$ is for the (1,0) ortho level.}
\end{table} 


\subsection{Heating and cooling mechanisms}

Since the discovery of excited metastable \hhhp~in the CMZ, several authors tried to determine the energy source that heats the gas at a few hundred Kelvin \citep{Yusef13}. Several hypothesis have been investigated: photoelectric effect, cosmic rays, X-rays, turbulence, or shocks.\\ 

\begin{figure*}[h!]
\begin{center}
\includegraphics[width=0.9\textwidth]{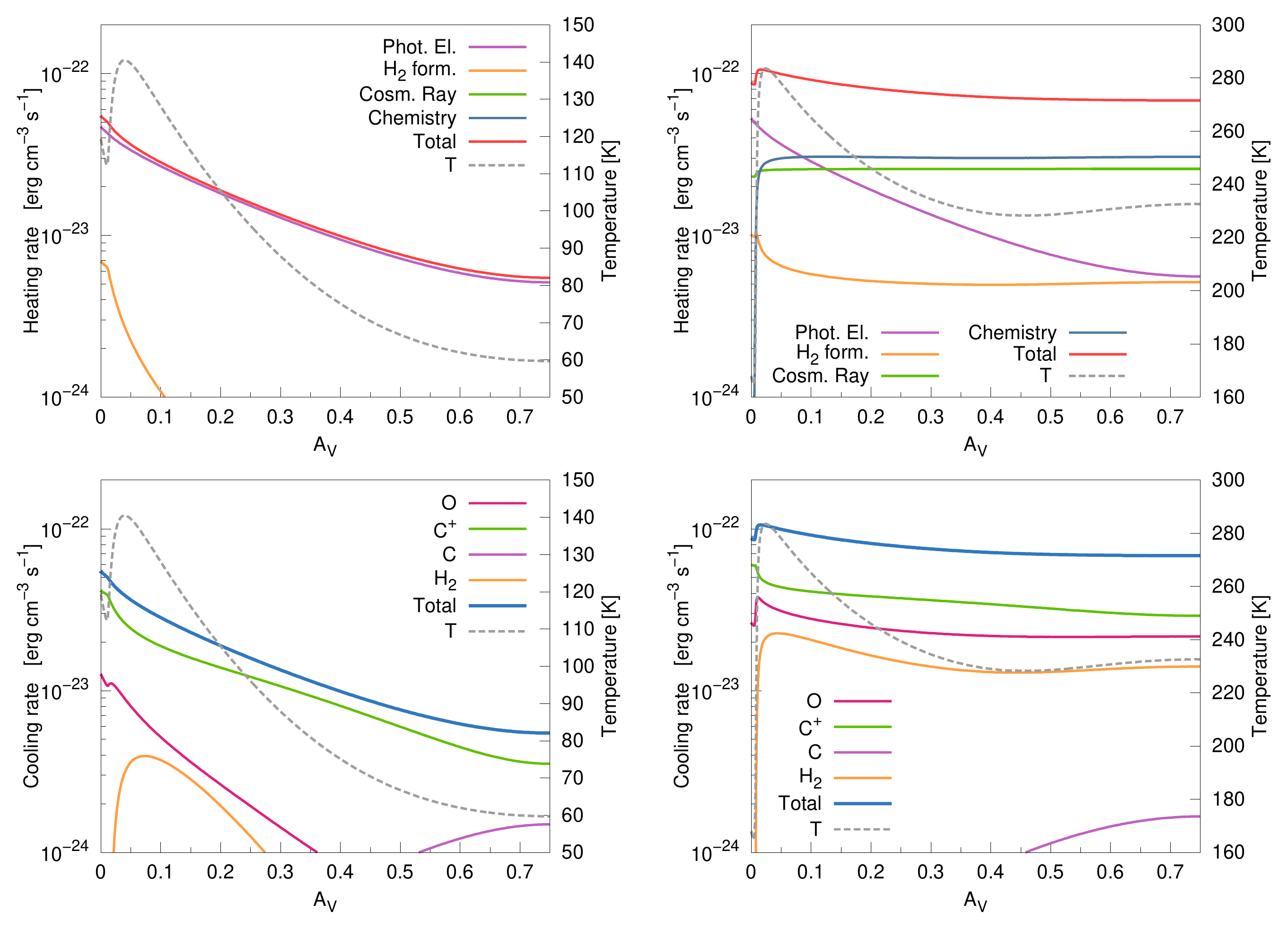}
\caption{Heating (top panels) and cooling (bottom panels) rates as a function of visual extinction for two models (left: $\zeta$ = 10$^{-16}$ s$^{-1}$; right: $\zeta = 7\times10^{-14}$ s$^{-1}$). In both models, \nh~=100 \cmt,  $G_0$ = 10 and, on both sides of the clouds, $A_\textrm{V}^{max}$ = 1.5, Z = 3. Figures present profiles from the edge of the cloud to the center. Cosmic ray heating and heating by exothermic reactions do not appear for the $\zeta$ = 10$^{-16}$ s$^{-1}$ model because their rate is always lower than $10^{-24}$ erg cm$^{-3}$ s$^{-1}$.}
\label{Fig:HeatCool}
\end{center}
\end{figure*}

Fig. \ref{Fig:HeatCool} shows gas temperature, heating, and cooling rates versus the position (in $A_\textrm{V}$) for two models with \nh~=~100 \cmt\  illuminated on both sides by a radiation field that is ten times higher than the ISRF, with a metallicity Z = 3 and with, for the first model, $\zeta = 1\times10^{-16}$ and for the second model, $\zeta = 7\times10^{-14}$ s$^{-1}$. We see that for $\zeta = 10^{-16}$ s$^{-1}$, the only efficient heating mechanism is the photoelectric effect on grains. The temperature does not exceed 140 K and is not sufficient to excite \hhhp~as observed in CMZ conditions. On the contrary, for $\zeta = 7\times10^{-14}$ s$^{-1}$, cosmic-ray ionizations and exothermic chemical reactions ($\Delta E = 4.7$ eV for the \hhhp~+ e$^{-}$ into the three H recombination channel) are efficient heating mechanisms through the whole cloud and keep the gas temperature at $T$ > 240 K. We find that our 4 eV injection per cosmic-ray H$_2$ ionization hypothesis is sufficient to heat the gas at temperatures around 250-300 K, consistent with \hhhp~excitation under CMZ conditions.\\

The major coolants are O and C$^+$ in the $\zeta = 10^{-16}$ s$^{-1}$ model. For the $\zeta = 7\times10^{-14}$ s$^{-1}$ model, cooling by H$_2$ starts to be important because of the high temperature. We find CO is not an important coolant in such diffuse gas because of a too low abundance.\\

\subsection{Limits and hypothesis of our models}

Our models have several limits. First, we use 1D isochoric PDR models. So, even though we account for detailed physical and chemical processes, the geometry is simplistic compared to reality. In particular, we assume only one \hhhp~component on each line of sight whereas HF absorption spectra show several components. For the J1747 line of sight, we tried a $\chi^2$ minimization on subcomponents assuming the same repartition of H$_3^+$ as HF. Cosmic-ray ionization rates deduced this way are on the same order of magnitude as that obtained considering only a single component.\\

An important factor of our models is the efficient H$_2$ formation rate on PAHs. Our PDR models reproduce $N(\textrm{H}_3^+)$ and its excitation at high $\zeta$ because we assume H atoms can efficiently form H$_2$ on PAHs via the Eley-Rideal mechanism. The formation rate of H$_2$ on PAHs is not well known, but some recent experiments suggest the process is efficient on the edge sites of PAHs \citep{Goumans11, Boshman12, Mennella12}. Moreover, ISO and Spitzer observations of H$_2$ emission lines in PDRs also suggest that H$_2$ must be formed efficiently on PAHs \citep{Habart04, Habart11}. If the H$_2$ formation rate was lower than that computed in our models, the deduced cosmic-ray flux would be lower (see Fig. \ref{Fig:EffectERPAH_NH3p}) and not efficient enough to heat the gas and explain \hhhp~excitation. Then, other heating sources would have to be found. We tested several other possibilities, and none of them are satisfying. Increasing the UV flux to heat the gas by photoelectric effect on grains is not a good hypothesis as the photodissociation rate of H$_2$ becomes too high, such that it becomes difficult to reproduce the observed column densities. We also investigated the role of X-rays heating with a new version of the Meudon PDR code (Godard et al., in preparation). Several X-rays sources can be found in the Galactic center. X-rays emitted by the ionized gas in the central cavity have a  flux that is too low to be responsible for the gas temperature in the CMZ (the most distant excited \hhhp~detected by \citealt{Goto11} is at more than 100 pc of Sgr A*). Another candidate could be heating by turbulent energy dissipation. Thanks to their vortex model, the TDR code, \cite{Godard14} suggested that intermittent turbulent energy dissipation may explain the presence of SH$^+$ and CH$^+$ observed by Herschel in the CMZ. Nevertheless, this process only heats  a small fraction of the gas and may not be efficient enough to explain the global heating of the CMZ. Introducing this scenario for \hhhp~would require a numerical model consistently taking   both PDR physics and chemistry into account as well as turbulent energy dissipation effects on chemistry and \hhhp~excitation. E. Bron \citep{EmericThese} developed a promising method to combine both approaches, based on the modeling of statistical properties of turbulence.\footnote{Available at https://tel.archives-ouvertes.fr/tel-01111148} We will  test this method in a future study.\\ 

\section{Conclusions}
\label{Sec:conclusion}

Our PDR models suggest a cosmic-ray ionization rate of H$_2$ in the CMZ of $\sim (1 - 11)\times10^{-14}$ s$^{-1}$.  The origin of this uncertainty range is due to the difficulty in constraining density from observations. This high value allows us to reproduce the observed column densities of \hhhp~and its unusual excitation in the metastable state (3,3) observed toward nine lines of sight in the CMZ spanning $\sim$100 pc apart from Sgr A*. The deduced cosmic-ray ionization rate is also compatible with the one deduced from synchrotron emission and Fe K$\alpha$ line emission \citep{Yusef13}. As a bonus, we also reproduce $N(\textrm{OH}^+)$, $N(\textrm{H}_2\textrm{O}^+)$ and $N(\textrm{H}_3\textrm{O}^+)$ observed in the CMZ by Herschel toward Sgr B2(N), a line of sight close to one line of sight where H$_3^+$ is detected. We confirm the suggestion by \cite{Oka04} that the gas probed by \hhhp~observations in the CMZ is diffuse and warm. Indeed, PDR models matching observations have a proton density $\lesssim$ 100 \cmt~and T$\sim$ 212-505 K. The main source of heating of the diffuse gas in the CMZ is the suprathermal electrons ejected from the gas by cosmic-ray ionizations and exothermic dissociative recombination reactions. If the analysis is only based on the J1747 line of sight, for which more constraints can be used, then models tend to favor gas densities that are lower than 50 \cmt. In that case, sizes of the diffuse components probed by \hhhp~are large (several ten parsecs to about 100 parsecs). This means that the diffuse gas probed by \hhhp~spans the whole CMZ. This is in agreement with \hhhp~observations that present the same behaviors in different directions in the CMZ.\\

We also show the importance of treating the detail of physical processes in astrochemical problems. We demonstrate how the simple analytic relationship between $N(\textrm{H}_3^+)$ and $\zeta$ (Eq. \ref{Eq:Indriolo}), commonly used to infer the cosmic-ray ionization rate in interstellar gas, may fail in an exotic medium. Indeed, for low and moderate cosmic-ray ionization rates the column density of \hhhp~increases linearly with $\zeta$ but decreases at high $\zeta$, in contradiction with the analytic expression. This is due to the ionization of H and H$_2$ by cosmic rays, which reduces the molecular fraction of the gas and increases the electronic fraction of the gas leading to fast \hhhp~recombination.\\

A key parameter of our results concerns the H$_2$ formation rate. We reached the conclusion that the H$_2$ formation rate in the diffuse gas of the CMZ must be enhanced when compared to local diffuse clouds. Indeed, in the CMZ, as in PDRs, the warm temperature of the gas opens the possibility of efficiently forming H$_2$ on grains and PAHs via chemisorption of H atoms followed by Eley-Rideal mechanism. Therefore, on one hand, a high cosmic-ray ionization rate reduces the number of available H atoms to form H$_2$ since they are ionized in H$^+$. On the other hand, heating the gas to several hundred Kelvin, cosmic rays open the efficient H$_2$ formation route via chemisorption sites. This allows us to reach high column densities of \hhhp~at large cosmic-ray ionization rates.\\ 

\begin{acknowledgements}
We thank Octavio Roncero who made  the H$_3^+$-H$_2$ collision rates available to us. We thank Takeshi Oka for useful discussions about \hhhp~and CO in the central molecular zone. Exchanges with Farhad Yusef-Zadeh and Vincent Tatischeff are acknowledged for their assistance in deriving  the cosmic-ray ionization rate from synchrotron emission. We also thank Micka\"el Coriat for his help concerning high energy astrophysics and radio counterpart as well as Thibaut Paumard for useful discussions about the Galactic center. This work was supported in part by grant SYMPATICO (ANR-11-BS56-0023) from the French Agence Nationale de la Recherche. It was also supported by the French CNRS national program PCMI. Maxime Ruaud acknowledges funding by the European Research Council (Starting Grant 3DICE, grant agreement 336474). 
\end{acknowledgements}

\bibliography{H3pGalCenter}

\begin{thebibliography}{89}
\expandafter\ifx\csname natexlab\endcsname\relax\def\natexlab#1{#1}\fi

\bibitem[{{Abia} {et~al.}(2015){Abia}, {Cunha}, {Cristallo}, \& {de
  Laverny}}]{Abia15}
{Abia}, C., {Cunha}, K., {Cristallo}, S., \& {de Laverny}, P. 2015, ArXiv
  e-prints

\bibitem[{{Aguado} {et~al.}(2010){Aguado}, {Barrag{\'a}n}, {Prosmiti},
  {Delgado-Barrio}, {Villarreal}, \& {Roncero}}]{Aguado10}
{Aguado}, A., {Barrag{\'a}n}, P., {Prosmiti}, R., {et~al.} 2010, \jcp, 133,
  024306

\bibitem[{{Bialy} \& {Sternberg}(2015)}]{Bialy15}
{Bialy}, S. \& {Sternberg}, A. 2015, \mnras, 450, 4424

\bibitem[{{Boschman} {et~al.}(2012){Boschman}, {Reitsma}, {Cazaux},
  {Schlath{\"o}lter}, {Hoekstra}, {Spaans}, \&
  {Gonz{\'a}lez-Maga{\~n}a}}]{Boshman12}
{Boschman}, L., {Reitsma}, G., {Cazaux}, S., {et~al.} 2012, \apjl, 761, L33

\bibitem[{{Brittain} {et~al.}(2004){Brittain}, {Simon}, {Kulesa}, \&
  {Rettig}}]{Brittain04}
{Brittain}, S.~D., {Simon}, T., {Kulesa}, C., \& {Rettig}, T.~W. 2004, \apj,
  606, 911

\bibitem[{{Bron}(2014)}]{EmericThese}
{Bron}, E. 2014, PhD thesis, {Université Paris Diderot}

\bibitem[{{Bron} {et~al.}(2014){Bron}, {Le Bourlot}, \& {Le Petit}}]{Bron14}
{Bron}, E., {Le Bourlot}, J., \& {Le Petit}, F. 2014, \aap, 569, A100

\bibitem[{{Compi{\`e}gne} {et~al.}(2011){Compi{\`e}gne}, {Verstraete}, {Jones},
  {Bernard}, {Boulanger}, {Flagey}, {Le Bourlot}, {Paradis}, \&
  {Ysard}}]{Compiegne11}
{Compi{\`e}gne}, M., {Verstraete}, L., {Jones}, A., {et~al.} 2011, \aap, 525,
  A103

\bibitem[{{Daflon} \& {Cunha}(2004)}]{Daflon04}
{Daflon}, S. \& {Cunha}, K. 2004, \apj, 617, 1115

\bibitem[{{Draine} \& {Li}(2007)}]{Draine07}
{Draine}, B.~T. \& {Li}, A. 2007, \apj, 657, 810

\bibitem[{{Faure} \& {Tennyson}(2003)}]{Faure03}
{Faure}, A. \& {Tennyson}, J. 2003, \mnras, 340, 468

\bibitem[{{Faure} {et~al.}(2006){Faure}, {Wiesenfeld}, {Valiron}, \& {et
  al.}}]{Faure06}
{Faure}, A., {Wiesenfeld}, L., {Valiron}, P., \& {et al.} 2006, Royal Society
  of London Philosophical Transactions Series A, 364, 3113

\bibitem[{{Federman} {et~al.}(1996){Federman}, {Weber}, \&
  {Lambert}}]{Federman96}
{Federman}, S.~R., {Weber}, J., \& {Lambert}, D.~L. 1996, \apj, 463, 181

\bibitem[{{Felenbok} \& {Roueff}(1996)}]{felenbok96}
{Felenbok}, P. \& {Roueff}, E. 1996, \apjl, 465, L57

\bibitem[{{Fitzpatrick} \& {Massa}(1986)}]{Fitzpatrick86}
{Fitzpatrick}, E.~L. \& {Massa}, D. 1986, \apj, 307, 286

\bibitem[{{Flower} \& {Pineau des Forets}(1998)}]{Flower98}
{Flower}, D.~R. \& {Pineau des Forets}, G. 1998, \mnras, 297, 1182

\bibitem[{{Geballe}(2012)}]{Geballe12}
{Geballe}, T.~R. 2012, Royal Society of London Philosophical Transactions
  Series A, 370, 5151

\bibitem[{{Geballe} {et~al.}(2006){Geballe}, {Goto}, {Usuda}, {Oka}, \&
  {McCall}}]{Geballe06}
{Geballe}, T.~R., {Goto}, M., {Usuda}, T., {Oka}, T., \& {McCall}, B.~J. 2006,
  \apj, 644, 907

\bibitem[{{Geballe} {et~al.}(1999){Geballe}, {McCall}, {Hinkle}, \&
  {Oka}}]{Geballe99}
{Geballe}, T.~R., {McCall}, B.~J., {Hinkle}, K.~H., \& {Oka}, T. 1999, \apj,
  510, 251

\bibitem[{{Geballe} \& {Oka}(1996)}]{Geballe96}
{Geballe}, T.~R. \& {Oka}, T. 1996, \nat, 384, 334

\bibitem[{{Geballe} \& {Oka}(2010)}]{Geballe10}
{Geballe}, T.~R. \& {Oka}, T. 2010, \apjl, 709, L70

\bibitem[{{Gerin} {et~al.}(2010{\natexlab{a}}){Gerin}, {de Luca}, {Black},
  {Goicoechea}, {Herbst}, {Neufeld}, {Falgarone}, {Godard}, {Pearson}, {Lis},
  {Phillips}, {Bell}, {Sonnentrucker}, {Boulanger}, {Cernicharo}, {Coutens},
  {Dartois}, {Encrenaz}, {Giesen}, {Goldsmith}, {Gupta}, {Gry}, {Hennebelle},
  {Hily-Blant}, {Joblin}, {Kazmierczak}, {Kolos}, {Krelowski},
  {Martin-Pintado}, {Monje}, {Mookerjea}, {Perault}, {Persson}, {Plume},
  {Rimmer}, {Salez}, {Schmidt}, {Stutzki}, {Teyssier}, {Vastel}, {Yu},
  {Contursi}, {Menten}, {Geballe}, {Schlemmer}, {Shipman}, {Tielens},
  {Philipp-May}, {Cros}, {Zmuidzinas}, {Samoska}, {Klein}, \&
  {Lorenzani}}]{Gerin10a}
{Gerin}, M., {de Luca}, M., {Black}, J., {et~al.} 2010{\natexlab{a}}, \aap,
  518, L110

\bibitem[{{Gerin} {et~al.}(2010{\natexlab{b}}){Gerin}, {de Luca}, {Goicoechea},
  {Herbst}, {Falgarone}, {Godard}, {Bell}, {Coutens}, {Ka{\'z}mierczak},
  {Sonnentrucker}, {Black}, {Neufeld}, {Phillips}, {Pearson}, {Rimmer},
  {Hassel}, {Lis}, {Vastel}, {Boulanger}, {Cernicharo}, {Dartois}, {Encrenaz},
  {Giesen}, {Goldsmith}, {Gupta}, {Gry}, {Hennebelle}, {Hily-Blant}, {Joblin},
  {Ko{\l}os}, {Kre{\l}owski}, {Mart{\'{\i}}n-Pintado}, {Monje}, {Mookerjea},
  {Perault}, {Persson}, {Plume}, {Salez}, {Schmidt}, {Stutzki}, {Teyssier},
  {Yu}, {Contursi}, {Menten}, {Geballe}, {Schlemmer}, {Morris}, {Hatch},
  {Imram}, {Ward}, {Caux}, {G{\"u}sten}, {Klein}, {Roelfsema}, {Dieleman},
  {Schieder}, {Honingh}, \& {Zmuidzinas}}]{Gerin10b}
{Gerin}, M., {de Luca}, M., {Goicoechea}, J.~R., {et~al.} 2010{\natexlab{b}},
  \aap, 521, L16

\bibitem[{{Gibb} {et~al.}(2010){Gibb}, {Brittain}, {Rettig}, {Troutman},
  {Simon}, \& {Kulesa}}]{Gibb10}
{Gibb}, E.~L., {Brittain}, S.~D., {Rettig}, T.~W., {et~al.} 2010, \apj, 715,
  757

\bibitem[{{Godard} {et~al.}(2012){Godard}, {Falgarone}, {Gerin}, {Lis}, {De
  Luca}, {Black}, {Goicoechea}, {Cernicharo}, {Neufeld}, {Menten}, \&
  {Emprechtinger}}]{Godard12}
{Godard}, B., {Falgarone}, E., {Gerin}, M., {et~al.} 2012, \aap, 540, A87

\bibitem[{{Godard} {et~al.}(2014){Godard}, {Falgarone}, \& {Pineau des
  For{\^e}ts}}]{Godard14}
{Godard}, B., {Falgarone}, E., \& {Pineau des For{\^e}ts}, G. 2014, \aap, 570,
  A27

\bibitem[{{G{\'o}mez-Carrasco} {et~al.}(2012){G{\'o}mez-Carrasco},
  {Gonz{\'a}lez-S{\'a}nchez}, {Aguado}, {Sanz-Sanz}, {Zanchet}, \&
  {Roncero}}]{Roncero12}
{G{\'o}mez-Carrasco}, S., {Gonz{\'a}lez-S{\'a}nchez}, L., {Aguado}, A.,
  {et~al.} 2012, \jcp, 137, 094303

\bibitem[{{Gonzalez Garcia} {et~al.}(2008){Gonzalez Garcia}, {Le Bourlot}, {Le
  Petit}, \& {Roueff}}]{Manuel08}
{Gonzalez Garcia}, M., {Le Bourlot}, J., {Le Petit}, F., \& {Roueff}, E. 2008,
  \aap, 485, 127

\bibitem[{{Goto} {et~al.}(2014){Goto}, {Geballe}, {Indriolo}, {Yusef-Zadeh},
  {Usuda}, {Henning}, \& {Oka}}]{Goto14}
{Goto}, M., {Geballe}, T.~R., {Indriolo}, N., {et~al.} 2014, \apj, 786, 96

\bibitem[{{Goto} {et~al.}(2013){Goto}, {Indriolo}, {Geballe}, \&
  {Usuda}}]{Goto13}
{Goto}, M., {Indriolo}, N., {Geballe}, T.~R., \& {Usuda}, T. 2013, Journal of
  Physical Chemistry A, 117, 9919

\bibitem[{{Goto} {et~al.}(2005){Goto}, {McCall}, {Geballe}, {Usuda}, \&
  {Oka}}]{Goto05}
{Goto}, M., {McCall}, B., {Geballe}, T., {Usuda}, T., \& {Oka}, T. 2005, in
  High Resolution Infrared Spectroscopy in Astronomy, ed. H.~U. {K{\"a}ufl},
  R.~{Siebenmorgen}, \& A.~{Moorwood}, 244--247

\bibitem[{{Goto} {et~al.}(2002){Goto}, {McCall}, {Geballe}, {Usuda},
  {Kobayashi}, {Terada}, \& {Oka}}]{Goto02}
{Goto}, M., {McCall}, B.~J., {Geballe}, T.~R., {et~al.} 2002, \pasj, 54, 951

\bibitem[{{Goto} {et~al.}(2011){Goto}, {Usuda}, {Geballe}, {Indriolo},
  {McCall}, {Henning}, \& {Oka}}]{Goto11}
{Goto}, M., {Usuda}, T., {Geballe}, T.~R., {et~al.} 2011, \pasj, 63, L13

\bibitem[{{Goto} {et~al.}(2008){Goto}, {Usuda}, {Nagata}, {Geballe}, {McCall},
  {Indriolo}, {Suto}, {Henning}, {Morong}, \& {Oka}}]{Goto08}
{Goto}, M., {Usuda}, T., {Nagata}, T., {et~al.} 2008, \apj, 688, 306

\bibitem[{{Goumans} \& {Bromley}(2011)}]{Goumans11}
{Goumans}, T.~P.~M. \& {Bromley}, S.~T. 2011, \mnras, 414, 1285

\bibitem[{{Grenier} {et~al.}(2015){Grenier}, {Black}, \& {Strong}}]{Grenier15}
{Grenier}, I.~A., {Black}, J.~H., \& {Strong}, A.~W. 2015, \araa, 53, 199

\bibitem[{{Gry} {et~al.}(2002){Gry}, {Boulanger}, {Nehm{\'e}}, {Pineau des
  For{\^e}ts}, {Habart}, \& {Falgarone}}]{Gry02}
{Gry}, C., {Boulanger}, F., {Nehm{\'e}}, C., {et~al.} 2002, \aap, 391, 675

\bibitem[{{Habart} {et~al.}(2011){Habart}, {Abergel}, {Boulanger}, {Joblin},
  {Verstraete}, {Compi{\`e}gne}, {Pineau Des For{\^e}ts}, \& {Le
  Bourlot}}]{Habart11}
{Habart}, E., {Abergel}, A., {Boulanger}, F., {et~al.} 2011, \aap, 527, A122

\bibitem[{{Habart} {et~al.}(2004){Habart}, {Boulanger}, {Verstraete},
  {Walmsley}, \& {Pineau des For{\^e}ts}}]{Habart04}
{Habart}, E., {Boulanger}, F., {Verstraete}, L., {Walmsley}, C.~M., \& {Pineau
  des For{\^e}ts}, G. 2004, \aap, 414, 531

\bibitem[{{Hartquist} {et~al.}(1978){Hartquist}, {Black}, \&
  {Dalgarno}}]{Hartquist78}
{Hartquist}, T.~W., {Black}, J.~H., \& {Dalgarno}, A. 1978, \mnras, 185, 643

\bibitem[{{Hugo} {et~al.}(2009){Hugo}, {Asvany}, \& {Schlemmer}}]{Hugo09}
{Hugo}, E., {Asvany}, O., \& {Schlemmer}, S. 2009, \jcp, 130, 164302

\bibitem[{{Indriolo} {et~al.}(2010){Indriolo}, {Blake}, {Goto}, {Usuda}, {Oka},
  {Geballe}, {Fields}, \& {McCall}}]{Indriolo10}
{Indriolo}, N., {Blake}, G.~A., {Goto}, M., {et~al.} 2010, \apj, 724, 1357

\bibitem[{{Indriolo} {et~al.}(2007){Indriolo}, {Geballe}, {Oka}, \&
  {McCall}}]{Indriolo07}
{Indriolo}, N., {Geballe}, T.~R., {Oka}, T., \& {McCall}, B.~J. 2007, \apj,
  671, 1736

\bibitem[{{Indriolo} \& {McCall}(2012)}]{Indriolo12}
{Indriolo}, N. \& {McCall}, B.~J. 2012, \apj, 745, 91

\bibitem[{{Indriolo} {et~al.}(2012){Indriolo}, {Neufeld}, {Gerin}, {Geballe},
  {Black}, {Menten}, \& {Goicoechea}}]{Indriolo12a}
{Indriolo}, N., {Neufeld}, D.~A., {Gerin}, M., {et~al.} 2012, \apj, 758, 83

\bibitem[{{Indriolo} {et~al.}(2015){Indriolo}, {Neufeld}, {Gerin}, {Schilke},
  {Benz}, {Winkel}, {Menten}, {Chambers}, {Black}, {Bruderer}, {Falgarone},
  {Godard}, {Goicoechea}, {Gupta}, {Lis}, {Ossenkopf}, {Persson},
  {Sonnentrucker}, {van der Tak}, {van Dishoeck}, {Wolfire}, \&
  {Wyrowski}}]{Indriolo15}
{Indriolo}, N., {Neufeld}, D.~A., {Gerin}, M., {et~al.} 2015, \apj, 800, 40

\bibitem[{{Jura}(1974)}]{Jura74}
{Jura}, M. 1974, \apj, 191, 375

\bibitem[{{Kokoouline} {et~al.}(2010){Kokoouline}, {Faure}, {Tennyson}, \&
  {Greene}}]{Kokoouline10}
{Kokoouline}, V., {Faure}, A., {Tennyson}, J., \& {Greene}, C.~H. 2010, \mnras,
  405, 1195

\bibitem[{{Le Bourlot} {et~al.}(2012){Le Bourlot}, {Le Petit}, {Pinto},
  {Roueff}, \& {Roy}}]{LeBourlot12}
{Le Bourlot}, J., {Le Petit}, F., {Pinto}, C., {Roueff}, E., \& {Roy}, F. 2012,
  \aap, 541, A76

\bibitem[{{Le Petit} {et~al.}(2009){Le Petit}, {Barzel}, {Biham}, {Roueff}, \&
  {Le Bourlot}}]{Pap2009}
{Le Petit}, F., {Barzel}, B., {Biham}, O., {Roueff}, E., \& {Le Bourlot}, J.
  2009, \aap, 505, 1153

\bibitem[{{Le Petit} {et~al.}(2006){Le Petit}, {Nehm{\'e}}, {Le Bourlot}, \&
  {Roueff}}]{Pap2006}
{Le Petit}, F., {Nehm{\'e}}, C., {Le Bourlot}, J., \& {Roueff}, E. 2006, \apjs,
  164, 506

\bibitem[{{Le Petit} {et~al.}(2004){Le Petit}, {Roueff}, \& {Herbst}}]{Pap2004}
{Le Petit}, F., {Roueff}, E., \& {Herbst}, E. 2004, \aap, 417, 993

\bibitem[{{Lindsay} \& {McCall}(2001)}]{Lindsay01}
{Lindsay}, C.~M. \& {McCall}, B.~J. 2001, Journal of Molecular Spectroscopy,
  210, 60

\bibitem[{{Lis} {et~al.}(2012){Lis}, {Schilke}, {Bergin}, \&
  {Emprechtinger}}]{Lis12}
{Lis}, D.~C., {Schilke}, P., {Bergin}, E.~A., \& {Emprechtinger}, M. 2012,
  Royal Society of London Philosophical Transactions Series A, 370, 5162

\bibitem[{{Lis} {et~al.}(2014){Lis}, {Schilke}, {Bergin}, {Gerin}, {Black},
  {Comito}, {De Luca}, {Godard}, {Higgins}, {Le Petit}, {Pearson},
  {Pellegrini}, {Phillips}, \& {Yu}}]{Lis14}
{Lis}, D.~C., {Schilke}, P., {Bergin}, E.~A., {et~al.} 2014, \apj, 785, 135

\bibitem[{{Liszt}(2003)}]{Liszt03}
{Liszt}, H. 2003, \aap, 398, 621

\bibitem[{{Liszt}(2006)}]{Liszt06}
{Liszt}, H.~S. 2006, Royal Society of London Philosophical Transactions Series
  A, 364, 3049

\bibitem[{{Mathis} {et~al.}(1983){Mathis}, {Mezger}, \& {Panagia}}]{Mathis83}
{Mathis}, J.~S., {Mezger}, P.~G., \& {Panagia}, N. 1983, \aap, 128, 212

\bibitem[{{Mathis} {et~al.}(1977){Mathis}, {Rumpl}, \& {Nordsieck}}]{Mathis77}
{Mathis}, J.~S., {Rumpl}, W., \& {Nordsieck}, K.~H. 1977, \apj, 217, 425

\bibitem[{{McCall} {et~al.}(1998){McCall}, {Geballe}, {Hinkle}, \&
  {Oka}}]{McCall98}
{McCall}, B.~J., {Geballe}, T.~R., {Hinkle}, K.~H., \& {Oka}, T. 1998, Science,
  279, 1910

\bibitem[{{McCall} {et~al.}(1999){McCall}, {Geballe}, {Hinkle}, \&
  {Oka}}]{McCall99}
{McCall}, B.~J., {Geballe}, T.~R., {Hinkle}, K.~H., \& {Oka}, T. 1999, \apj,
  522, 338

\bibitem[{{McCall} {et~al.}(2002){McCall}, {Hinkle}, {Geballe},
  {Moriarty-Schieven}, {Evans}, {Kawaguchi}, {Takano}, {Smith}, \&
  {Oka}}]{McCall02}
{McCall}, B.~J., {Hinkle}, K.~H., {Geballe}, T.~R., {et~al.} 2002, \apj, 567,
  391

\bibitem[{{McCall} {et~al.}(2003){McCall}, {Huneycutt}, {Saykally}, {Geballe},
  {Djuric}, {Dunn}, {Semaniak}, {Novotny}, {Al-Khalili}, {Ehlerding},
  {Hellberg}, {Kalhori}, {Neau}, {Thomas}, {{\"O}sterdahl}, \&
  {Larsson}}]{McCall03}
{McCall}, B.~J., {Huneycutt}, A.~J., {Saykally}, R.~J., {et~al.} 2003, \nat,
  422, 500

\bibitem[{{Mennella} {et~al.}(2012){Mennella}, {Hornek{\ae}r}, {Thrower}, \&
  {Accolla}}]{Mennella12}
{Mennella}, V., {Hornek{\ae}r}, L., {Thrower}, J., \& {Accolla}, M. 2012,
  \apjl, 745, L2

\bibitem[{{Meyer} {et~al.}(1997){Meyer}, {Cardelli}, \& {Sofia}}]{Meyer97}
{Meyer}, D.~M., {Cardelli}, J.~A., \& {Sofia}, U.~J. 1997, \apjl, 490, L103

\bibitem[{{Meyer} {et~al.}(1998){Meyer}, {Jura}, \& {Cardelli}}]{Meyer98}
{Meyer}, D.~M., {Jura}, M., \& {Cardelli}, J.~A. 1998, \apj, 493, 222

\bibitem[{{Moskalenko} {et~al.}(2006){Moskalenko}, {Porter}, \&
  {Strong}}]{Moskalenko06}
{Moskalenko}, I.~V., {Porter}, T.~A., \& {Strong}, A.~W. 2006, \apjl, 640, L155

\bibitem[{{Neufeld} {et~al.}(2010){Neufeld}, {Goicoechea}, {Sonnentrucker},
  {Black}, {Pearson}, {Yu}, {Phillips}, {Lis}, {de Luca}, {Herbst}, {Rimmer},
  {Gerin}, {Bell}, {Boulanger}, {Cernicharo}, {Coutens}, {Dartois},
  {Kazmierczak}, {Encrenaz}, {Falgarone}, {Geballe}, {Giesen}, {Godard},
  {Goldsmith}, {Gry}, {Gupta}, {Hennebelle}, {Hily-Blant}, {Joblin},
  {Ko{\l}os}, {Kre{\l}owski}, {Mart{\'{\i}}n-Pintado}, {Menten}, {Monje},
  {Mookerjea}, {Perault}, {Persson}, {Plume}, {Salez}, {Schlemmer}, {Schmidt},
  {Stutzki}, {Teyssier}, {Vastel}, {Cros}, {Klein}, {Lorenzani}, {Philipp},
  {Samoska}, {Shipman}, {Tielens}, {Szczerba}, \& {Zmuidzinas}}]{Neufeld10}
{Neufeld}, D.~A., {Goicoechea}, J.~R., {Sonnentrucker}, P., {et~al.} 2010,
  \aap, 521, L10

\bibitem[{{Neufeld} {et~al.}(2005){Neufeld}, {Wolfire}, \&
  {Schilke}}]{Neufeld05}
{Neufeld}, D.~A., {Wolfire}, M.~G., \& {Schilke}, P. 2005, \apj, 628, 260

\bibitem[{{Oka}(2013)}]{Oka13}
{Oka}, T. 2013, Chemical Reviews, 113, 8738

\bibitem[{{Oka}(2015)}]{Oka15}
{Oka}, T. 2015, in American Institute of Physics Conference Series, Vol. 1642,
  American Institute of Physics Conference Series, 373--376

\bibitem[{{Oka} \& {Epp}(2004)}]{Oka04}
{Oka}, T. \& {Epp}, E. 2004, \apj, 613, 349

\bibitem[{{Oka} {et~al.}(2005){Oka}, {Geballe}, {Goto}, {Usuda}, \&
  {McCall}}]{Oka05}
{Oka}, T., {Geballe}, T.~R., {Goto}, M., {Usuda}, T., \& {McCall}, B.~J. 2005,
  \apj, 632, 882

\bibitem[{{Oka} {et~al.}(2012){Oka}, {Onodera}, {Nagai}, {Tanaka}, {Matsumura},
  \& {Kamegai}}]{Oka12}
{Oka}, T., {Onodera}, Y., {Nagai}, M., {et~al.} 2012, \apjs, 201, 14

\bibitem[{{Park} \& {Light}(2007)}]{Park07}
{Park}, K. \& {Light}, J.~C. 2007, \jcp, 126, 044305

\bibitem[{{Porter} \& {Strong}(2005)}]{Porter05}
{Porter}, T.~A. \& {Strong}, A.~W. 2005, International Cosmic Ray Conference,
  4, 77

\bibitem[{{Rachford} {et~al.}(2002){Rachford}, {Snow}, {Tumlinson}, {Shull},
  {Blair}, {Ferlet}, {Friedman}, {Gry}, {Jenkins}, {Morton}, {Savage},
  {Sonnentrucker}, {Vidal-Madjar}, {Welty}, \& {York}}]{Rachford02}
{Rachford}, B.~L., {Snow}, T.~P., {Tumlinson}, J., {et~al.} 2002, \apj, 577,
  221

\bibitem[{{Rolleston} {et~al.}(2000){Rolleston}, {Smartt}, {Dufton}, \&
  {Ryans}}]{Rolleston00}
{Rolleston}, W.~R.~J., {Smartt}, S.~J., {Dufton}, P.~L., \& {Ryans}, R.~S.~I.
  2000, \aap, 363, 537

\bibitem[{{Roueff}(1996)}]{roueff96}
{Roueff}, E. 1996, \mnras, 279, L37

\bibitem[{{Rudolph} {et~al.}(2006){Rudolph}, {Fich}, {Bell}, {Norsen},
  {Simpson}, {Haas}, \& {Erickson}}]{Rudolph06}
{Rudolph}, A.~L., {Fich}, M., {Bell}, G.~R., {et~al.} 2006, \apjs, 162, 346

\bibitem[{{Savage} \& {Sembach}(1996)}]{Savage96}
{Savage}, B.~D. \& {Sembach}, K.~R. 1996, \araa, 34, 279

\bibitem[{{Schilke} {et~al.}(2010){Schilke}, {Comito}, {M{\"u}ller}, {Bergin},
  {Herbst}, {Lis}, {Neufeld}, {Phillips}, {Bell}, {Blake}, {Cabrit}, {Caux},
  {Ceccarelli}, {Cernicharo}, {Crockett}, {Daniel}, {Dubernet},
  {Emprechtinger}, {Encrenaz}, {Falgarone}, {Gerin}, {Giesen}, {Goicoechea},
  {Goldsmith}, {Gupta}, {Joblin}, {Johnstone}, {Langer}, {Latter}, {Lord},
  {Maret}, {Martin}, {Melnick}, {Menten}, {Morris}, {Murphy}, {Ossenkopf},
  {Pagani}, {Pearson}, {P{\'e}rault}, {Plume}, {Qin}, {Salez}, {Schlemmer},
  {Stutzki}, {Trappe}, {van der Tak}, {Vastel}, {Wang}, {Yorke}, {Yu},
  {Erickson}, {Maiwald}, {Kooi}, {Karpov}, {Zmuidzinas}, {Boogert}, {Schieder},
  \& {Zaal}}]{Schilke10}
{Schilke}, P., {Comito}, C., {M{\"u}ller}, H.~S.~P., {et~al.} 2010, \aap, 521,
  L11

\bibitem[{{Schilke} {et~al.}(2013){Schilke}, {Lis}, {Bergin}, {Higgins}, \&
  {Comito}}]{Schilke13}
{Schilke}, P., {Lis}, D.~C., {Bergin}, E.~A., {Higgins}, R., \& {Comito}, C.
  2013, Journal of Physical Chemistry A, 117, 9766

\bibitem[{{Seaton}(1979)}]{Seaton79}
{Seaton}, M.~J. 1979, \mnras, 187, 73P

\bibitem[{{Snow} {et~al.}(2007){Snow}, {Destree}, \& {Jensen}}]{Snow07}
{Snow}, T.~P., {Destree}, J.~D., \& {Jensen}, A.~G. 2007, \apj, 655, 285

\bibitem[{{Sternberg} {et~al.}(2014){Sternberg}, {Le Petit}, {Roueff}, \& {Le
  Bourlot}}]{Sternberg14}
{Sternberg}, A., {Le Petit}, F., {Roueff}, E., \& {Le Bourlot}, J. 2014, \apj,
  790, 10

\bibitem[{{Strong} {et~al.}(2010){Strong}, {Porter}, {Digel},
  {J{\'o}hannesson}, {Martin}, {Moskalenko}, {Murphy}, \& {Orlando}}]{Strong10}
{Strong}, A.~W., {Porter}, T.~A., {Digel}, S.~W., {et~al.} 2010, \apjl, 722,
  L58

\bibitem[{{Wakelam} {et~al.}(2012){Wakelam}, {Herbst}, {Loison}, {Smith},
  {Chandrasekaran}, {Pavone}, {Adams}, {Bacchus-Montabonel}, {Bergeat},
  {B{\'e}roff}, {Bierbaum}, {Chabot}, {Dalgarno}, {van Dishoeck}, {Faure},
  {Geppert}, {Gerlich}, {Galli}, {H{\'e}brard}, {Hersant}, {Hickson},
  {Honvault}, {Klippenstein}, {Le Picard}, {Nyman}, {Pernot}, {Schlemmer},
  {Selsis}, {Sims}, {Talbi}, {Tennyson}, {Troe}, {Wester}, \&
  {Wiesenfeld}}]{Wakelam12}
{Wakelam}, V., {Herbst}, E., {Loison}, J.-C., {et~al.} 2012, \apjs, 199, 21

\bibitem[{{Yusef-Zadeh} {et~al.}(2013){Yusef-Zadeh}, {Hewitt}, {Wardle},
  {Tatischeff}, {Roberts}, {Cotton}, {Uchiyama}, {Nobukawa}, {Tsuru}, {Heinke},
  \& {Royster}}]{Yusef13}
{Yusef-Zadeh}, F., {Hewitt}, J.~W., {Wardle}, M., {et~al.} 2013, \apj, 762, 33

\end{thebibliography}
\bibliographystyle{aa}

\end{document}